\documentclass[showpacs, pra,onecolumn,preprintnumbers ,amsmath, amssymb, superscriptaddress, aps]{revtex4-2}
\usepackage{color}
\usepackage{amsmath,amssymb}
\usepackage{pifont}
\usepackage{amssymb}  
\usepackage{bbold}
\usepackage{float}
\usepackage{subfloat}

\usepackage[caption=false]{subfig}
\usepackage{tikz}
\usepackage{makecell}
\usepackage{subfig}
\usepackage{pifont}   
\usepackage{graphicx} 
\graphicspath{{Figures/}}
\usepackage{dcolumn}  
\usepackage{bm}       
\usepackage{multirow} 
\usepackage{placeins}
\usepackage[colorlinks]{hyperref}
\usepackage{mathtools}
\usepackage{appendix}

\newcommand{\ket}[1]{\left|{#1}\right\rangle}

\begin{document}
	
	\title{Virtual excitations and entanglement dynamics and polygamy \\ in three ultra-strongly coupled systems
	}
	\date{\today}
	\author{Radouan Hab-arrih}
	\affiliation{Laboratory of Theoretical Physics, Faculty of Sciences, Choua\"ib Doukkali University, PO Box 20, 24000 El Jadida, Morocco}
	
	\author{Ahmed Jellal}
	\email{a.jellal@ucd.ac.ma}
	\affiliation{Laboratory of Theoretical Physics, Faculty of Sciences, Choua\"ib Doukkali University, PO Box 20, 24000 El Jadida, Morocco}
	\affiliation{Canadian Quantum  Research Center,
		204-3002 32 Ave Vernon,  BC V1T 2L7,  Canada}
	

			

			\begin{abstract}
				
The Milburn dynamics of three non resonant ultra-strongly coupled oscillators are resolved by using symplectic geometry. We look at the Milburn dynamics of virtual excitations and how they 
affect the pairwise entanglement. It is found that the dynamics of excitations and entanglement experience similar profiles against time, physical parameters, and decoherence rate. Furthermore, we show that the extinction of excitations entails separability, which demonstrates the hierarchy between entanglement and virtual excitations.  Additionally, we analyze the effects of physical parameters on the redistribution of virtual excitations among the three bi-partitions. As a result, we show the violation of the monogamy of excitations as in quantum discord. This implies that excitations can be considered  as signatures of quantum correlations beyond entanglement. Besides, we emphasize that our treatment can be used to model coupled quantum circuits in  real situations (with decoherence).  
				
			\end{abstract}
			
			\maketitle

\section{Introduction}
Ultra strong coupling (USC) physics, in which the coupling becomes comparable to the system's frequencies, has emerged as a new and increasingly popular domain \cite{NoriRev,USC1, USC2, USC3, USC4, USC5, USC6}. The importance of the USC lies in its potential effects on several physical phenomena, including photon blockade \cite{photonblockade}, Purcell effect \cite{PE}, Zeno effect \cite{Zeno}, and maintaining entanglement by virtual excitations of a  vacuum state \cite{Rad2021oct,maintaining}.  The fluctuations of the vacuum are one of the most striking phenomena of modern physics and have colossal applications in quantum technologies \cite{vac1, vac2}. The vacuum in USC systems becomes populated by  short-lived particles called virtual (not being absorbed by detectors) excitations \cite{NoriRev}. These excitations are responsible for mediating forces (i.e., the Casimir effect) \cite{Casimir} and the transfer of energy between oscillators and atoms \cite{transfer}. Moreover, we point out that the known counter-rotating terms (CRT) are responsible for simultaneous creation and annihilation, as well as the amplification of excitations in a vacuum state. As a result, the rotating wave approximation (RWA), where CRT is neglected, breaks down \cite{WRA, USC1, USC6}. 

Multipartite systems made of coupled oscillators are central in physics \cite{multi1,multi3,multi4} and for reviews we cite the references \cite{Walter2016,Oliveira2017,chiara2018}. In fact, they lead to the modeling of coupled ions in ion traps \cite{trap}, arrayed coupled nano-sized electromechanical devices  \cite{Nano}, light propagation in inhomogeneous media \cite{light}, and a nitrogen vacancy ensemble embedded in a diamond nanobeam \cite{nano}. Besides, the separability of a tripartite gaussian state was studied in \cite{separability}. The stationary and non-stationary entanglement of three oscillators were exhaustively studied in \cite{Merdaci2020, Rad2021}, respectively. Moreover, avoiding external decoherence and disentanglement in oscillator systems was studied in \cite{avoiding, avoiding1}. The interplay between coherence and entanglement and their redistribution was discussed in \cite{Rad2021, Redist}. Additionally, it was shown that with three coupled parametric oscillators, it is possible to generate entanglement even at high temperatures \cite{thermal}. Using the optimal control theory, maximizing entanglement in coupled oscillators is discussed by Stefanatos in \cite{stefanatos}.

Motivated by our last achievement with the Milburn dynamics (MD) of two ultra-strongly coupled oscillators \cite{milburn2d}, showing the interplay between steering and entanglement with virtual excitations, we seek to study the MD for three oscillators. The current paper surveys the Milburn dynamics of three ultra-strongly coupled oscillators. We harness the covariance matrix formalism beyond RWA. Thus,  the interconnection of excitations with pairwise entanglement will be legitimately discussed in USC.  Additionally, this work sheds light on whether excitations are monogamous or polygamous. The monogamy constraints are fundamental in quantum information science \cite{monogamy} because they capture the quantumness of correlation. The entanglement, in particular, is monogamous, i.e., $E(a|b)+E(a|c)\leq E(a|bc)$, with $a, b$ and $c$ are representing three distinct parties. Nevertheless, it is not the case for quantum discord \cite{violation}. Furthermore, we ask whether excitations are monogamous or not. Addressing such questions of monogamy is important to understanding the role of virtual excitations as signatures of genuine non-classical correlations beyond entanglement.

The present paper is formulated as follows. In Sec.  \ref{sec2}, we define the system made of three oscillators as well as present the diagonalization scheme. While, in Sec. \ref{sec 3}, we write out the unitary  transformations in their symplectic form and  resolve the Milburn dynamics beyond the RWA. Sec. \ref{sec 4} investigates quantum entanglement and virtual excitations.  In Sec. \ref{sec 5}, we present our numerical results together with some discussions. Finally, we conclude our work.

\section{Model and Diagonalization \label{sec2} }

\begin{figure}[htbphtbp]
	\centering
	\includegraphics[width=0.6\linewidth]{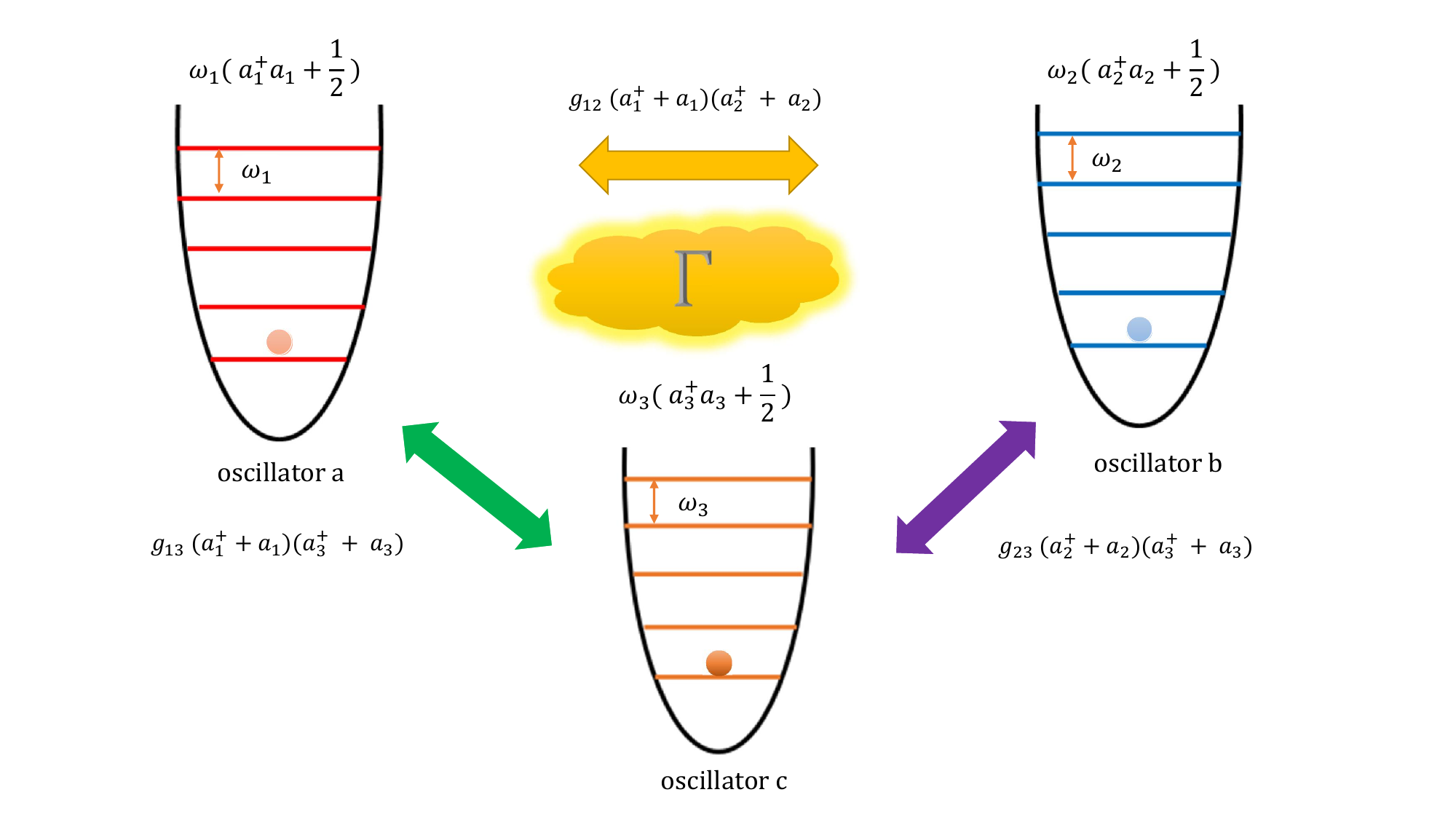}
	\caption{(color online)
		Schematic diagram of three coupled harmonic oscillators  initially prepared in $ \ket{000} $. Each of the two oscillators $ j  $ and $ k $ are coupled via "position-position" $ x_{j}x_{k} $ type interaction with coupling strength $ g_{jk} $, $ (j<k) $. The dynamics of the system is affected by the intrinsic decoherence quantified by the Milburn decoherence parameter $\Gamma$. }\label{fig0}
\end{figure}

We consider three interacting harmonic oscillators $ (a,b,c) $,  in which each oscillator has its own angular frequency $ \omega_{i} $ ($ i=1,2,3 $), as shown in Fig. \ref{fig0}. The interaction between  two oscillators $ k $ and $ l $ 
driven by a time-independent constant $ J_{kl} $, with $ 1\leq k<l\leq 3 $. 

The total Hamiltonian describing the isolated system is
\begin{eqnarray}
H=\frac{p_{1}^{2}}{2}+\frac{p_{2}^{2}}{2}+\frac{p_{3}^{2}}{2}+\frac{1}{2}\omega_{1}^{2}x_{1}^{2}+\frac{1}{2}\omega_{2}^{2}x_{2}^{2}+\frac{1}{2}\omega_{3}^{2}x_{3}^{2}+J_{12}x_{1}x_{2}+J_{13}x_{1}x_{3}+J_{23}x_{2}x_{3} \label{eq1}
\end{eqnarray} 
where the position $ x_{i} $ and momentum $ p_{i} $ operators are conjugate, satisfying the  commutation relations $ \left[x_{k},p_{l}\right] =i \delta_{kl} $ and $ \left[x_{k},x_{l}\right] =\left[p_{k},p_{l}\right]=0$. For instance, this Hamiltonian can be used to describe three capacitively coupled $LC$ oscillators \cite{multi1}. 
The masses are set to be equal at 1, and a simple transformation validates the assumption \cite{macedo}, and setting $\hbar=1$. In matrix form, we have 
\begin{eqnarray}
  H&=& \frac{1}{2}\mathbb{P}^{T}\mathbb{P}+\frac{1}{2} \mathbb{X}^{T}\mathbb{V}\mathbb{X}
\end{eqnarray}
 where 
 $\mathbb{P}=(p_1,p_2,p_3)^{T}$, $\mathbb{X}=(x_1,x_2,x_3)^{T}$, with ${T}$ stands for the transpose, and  $\mathbb{V}$ is the  potential matrix given by
 \begin{eqnarray}
 \mathbb{V}=\begin{pmatrix}
 \omega_{1}^{2} & J_{12} & J_{13} \\
 J_{12} & \omega_{2}^{2} & J_{23} \\
 J_{13} & J_{23} & \omega_{3}^{2}
 \end{pmatrix}.
 \end{eqnarray}
Using the Sylvester criterion of the positivity of $\mathbb{V}$, we obtain a manifold of physical parameters that gives a bound state. This is
 \begin{eqnarray}
 \mathcal{B}=\lbrace  (\omega_{j}, J_{jk})_{1\leq j<k\leq3}\in \mathbb{C}^{6}, \quad \max(\omega_{1}^{2},0) \times\max(\omega_{2}^{2}\omega_{3}^2-J_{23}^{2},0)\times \max(\det(\mathbb{V}),0)>0 \rbrace.
 \end{eqnarray}
Now, to diagonalize the Hamiltonian using a unitary transformation, we address the time-independent Euler unitary transformation $ \mathbb{T} (\alpha, \beta, \gamma) $  defined as \cite{Merdaci2020,Rad2021, lohe, multi5,multi6}
\begin{eqnarray}
\mathbb{T} (\alpha, \beta, \gamma):= e^{-i \alpha \mathcal{J}_{3}}\times e^{-i \beta \mathcal{J}_{2}}\times e^{-i \gamma \mathcal{J}_{3}}\label{rotations}
 \end{eqnarray} 
where $\mathcal{J}_{2,3}$ are the $y$ and $z$ components of the angular momentum operator $\mathcal{J}$, respectively. By making the following choices of the Euler angles, $ \alpha, \beta $ and $\gamma$ \cite{{Rad2021}}
 \begin{align}
 &
 \cos(2\alpha)=2\left[\frac{\left(\Omega_{2}^{4}-(\omega_{1}^{2}+\omega_{2}^{2})\Omega_{2}^{2}+\omega_{1}^{2}\omega_{2}^{2}-J_{12}^{2}\right)\frac{\left(\Omega_{1}^{2}-\Omega_{3}^{2}\right)}{\left(\Omega_{3}^{2}-\Omega_{2}^{2}\right)}}{\left(\Omega_{1}^{4}-(\omega_{1}^{2}+\omega_{2}^{2})\Omega_{1}^{2}+\omega_{1}^{2}\omega_{2}^{2}-J_{12}^{2}\right)}+1\right]^{-1}-1\\\ 
 &
 \cos(2\beta)=2\frac{\Omega_{3}^{4}-(\omega_{1}^{2}+\omega_{2}^{2})\Omega_{3}^{2}+\omega_{2}^{2}\omega_{1}^{2}-J_{12}^{2}}{(\Omega_{3}^{2}-\Omega_{1}^{2})(\Omega_{3}^{2}-\Omega_{2}^{2})}-1\\
 &
 \cos(2\gamma)=2\left[ \frac{\Omega_{3}^{4}-(\omega_{1}^{2}+\omega_{3}^{2})\Omega_{3}^{2}+\omega_{1}^{2}\omega_{3}^{2}-J_{13}^{2} }{\Omega_{3}^{4}-(\omega_{2}^{2}+\omega_{3}^{2})\Omega_{3}^{2}+\omega_{2}^{2}\omega_{3}^{2}-J_{23}^{2} }+1\right]^{-1}-1
 \end{align} 
 we end up with a transformed Hamiltonian
 \begin{align}\label{diaH}
 H_{D}
 = \frac{1}{2}\mathbb{P}^{t}\mathbb{P}+\frac{1}{2}{\mathbb{X}}^{t}\mathbb{V}_{D}\mathbb{X}
 \end{align}
including the  diagonalized potential matrix 
	$\mathbb{V}_{D}=\text{diag}\left(\Omega_{1}^{2},\Omega_{2}^{2},\Omega_{3}^{2}\right)$
 and the normal  frequencies \cite{{Rad2021}}
 \begin{eqnarray}
 \Omega_{1}^{2}=\frac{1}{3}\left[\varpi+2\sqrt{p}\cos(\Phi)\right],\quad
 \Omega_{2}^{2}=\frac{1}{3}\left[\varpi+2\sqrt{p}\cos\left(\Phi+\frac{2\pi}{3}\right)\right],\quad
 \Omega_{3}^{2}=\frac{1}{3}\left[\varpi+2\sqrt{p}\cos\left(\Phi-\frac{2\pi}{3}
 \right)\right]
 \end{eqnarray}
where the  parameters  $\varpi$, $p$ and $\Phi$  read as
 \begin{align}
& c_{0}=\sum\limits_{(i,i)\neq(j,k),j<k}^{3}\omega^{2}_{i}J_{jk}^{2}-\prod\limits_{i=1}^{3}\omega^{2}_{i}+2\prod\limits_{i<j}^{3}J_{ij},\quad \varpi=\sum\limits_{i=1}^{3}\omega^{2}_{i},\quad 
 c_{1}=\sum\limits_{i<j}^{3} \omega_{i}^{2}\omega_{j}^{2}-\sum\limits_{i<j}^{3}J_{ij}^{2}\\
 &
  \Phi=\frac{1}{3} \arctan\left(\frac{\sqrt{p^{3}-q^{2}}}{q}\right), \quad 
p= \varpi^{2}-3c_{1},\quad q=-\frac{27}{2}c_{0}+\varpi^{3}-\frac{9}{2} c_{1}\varpi.
 \end{align}
As a result,  we can write \eqref{diaH} in the following diagonal form
\begin{eqnarray}
H_{D}&=&\frac{p_{1}^{2}}{2}+\frac{p_{2}^{2}}{2}+\frac{p_{3}^{2}}{2}+\frac{1}{2}\Omega_{1}^{2}x_{1}^{2}+\frac{1}{2}\Omega_{2}^{2}x_{2}^{2}+\frac{1}{2}\Omega_{3}^{2}x_{3}^{2}.
\end{eqnarray}

To discuss entanglement together with virtual excitations, we employ the annihilation {$ \hat{a}_{j} $ and creation   $\hat{a}^{\dagger}_{j} $ operators via the mapping 
\begin{align}
	\hat{a}_{j}=\left( \hat{a}_{j}^{\dagger}\right)^{\dagger}  =\frac{1}{\sqrt{2\omega_{j}}}\left(\omega_{j}x_{j}+ip_{j}\right), \quad j=1,2,3.
\end{align}
These enable us to cast the non-diagonalized Hamiltonian (\ref{eq1}) as 
\begin{eqnarray}
 H=\sum\limits_{j=1}^{3}\omega_{j}\left(\hat{a}^{\dagger}_{j}\hat{a}_{j}+\frac{1}{2}\right)+\sum\limits_{j<k}^{3} g_{jk}(\hat{a}_{j}^{\dagger}+\hat{a}_{j})(\hat{a}^{\dagger}_{k}+\hat{a}_{k})
 \end{eqnarray}
 where the coupling strengths have been set 
$ g_{jk}= \frac{J_{jk}}{2\sqrt{\omega_{j}\omega_{k}}}, 1\leq j<k\leq 3$.
Also, the previous Euler transformation (\ref{rotations}) leads to a diagonalized Hamiltonian 
\begin{eqnarray}
H_{D}&=&\sum\limits_{j=1}^{3}\Omega_{j}\left(\mathfrak{A}^{+}_{j}\mathfrak{A}_{j}+\frac{1}{2}\right)\label{diagonal}
\end{eqnarray}
in terms of the operators 
\begin{align}
\mathfrak{A}_{j}=\left( \mathfrak{A}_{j}^{+}\right)^{+}= \frac{1}{\sqrt{2\Omega_{j}}}\left(\Omega_{j}x_{j}+ip_{j}\right) 	
\end{align}
which satisfy the commutation relation  $ \left[\mathfrak{A}_{k},\mathfrak{A}^{+}_{l} \right]=\delta_{kl}  $, and the rest are null. 
 As a result, the Hamiltonian  in terms of the diagonal operator  reads as 
\begin{eqnarray}
H= \mathbb{T}^{-1} (\alpha, \beta, \gamma) H_{D} \mathbb{T} (\alpha, \beta, \gamma)
\end{eqnarray} 
or equivalently
\begin{eqnarray}
H=e^{i\gamma \mathcal{J}_{3}}\ e^{i\beta \mathcal{J}_{2}}\ e^{i\alpha \mathcal{J}_{3}}\ H_{D}\ e^{-i\alpha \mathcal{J}_{3}}\ e^{-i\beta \mathcal{J}_{2}}\ e^{-i\gamma \mathcal{J}_{3}}\label{eq20 }.
\end{eqnarray}
Now, it is worthwhile to note that  $ H_{D} $ is diagonal in the representation $ \left\lbrace \mathfrak{A}_{j},  \mathfrak{A}_{j}^{+} \right\rbrace $. To diagonalize the Hamiltonian in terms of original  operators {$\hat{a}_{j}$ and $\hat{a}_{j}^{\dagger}$,  we rewrite (\ref{diagonal}) as
\begin{eqnarray}
 H_{D}=\sum\limits_{j=1}^{3}\tilde{\Omega}_{j}\left(\hat{a}_{j}^{\dagger}\hat{a}_{j}+\frac{1}{2}\right)+\mathfrak{g}_{j}\left(\left(\hat{a}_{j}^{\dagger}\right)^{2}+\hat{a}_{j}^{2}\right)
\end{eqnarray}
where we have defined 
$\tilde{\Omega}_{j}=\frac{1}{2}\left( \frac{\Omega_{j}^{2}}{\omega_{j}}+\omega_{j}\right), \mathfrak{g}_{j}=\frac{1}{4}\left( \frac{\Omega_{j}^{2}}{\omega_{j}}-\omega_{j}\right)$
and  perform the following squeezing operators 
\begin{eqnarray}
S_{j}(r_{j})=e^{\frac{r_{j}}{2}\left[ \hat{a}_{j}^{2}-(\hat{a}_{j}^{\dagger})^{2}\right] }, \quad r_{j}=\frac{1}{2}\ln\left( \frac{\Omega_{j}}{\omega_{j}}\right) 
\end{eqnarray}
giving rise to the  following diagonal Hamiltonian in the original basis
\begin{eqnarray}
H_{d}= \prod\limits_{j=1}^{3}S_{j}(-r_{j}) \ H_{D}\ \prod\limits_{j=1}^{3}S_{j}(r_{j})= \sum\limits_{j=1}^{3}\Omega_{j}\left(\hat{a}_{j}^{\dagger}\hat{a}_{j}+\frac{1}{2}\right).
\end{eqnarray} 
Combining all to map the Hamiltonian \eqref{eq1} as
\begin{eqnarray}
H&=& e^{i\gamma \mathcal{J}_{3}}\ e^{i\beta \mathcal{J}_{2}}\ e^{i\alpha \mathcal{J}_{3}}\ S_{1}(-r_{1})\ S_{2}(-r_{2})\ S_{3}(-r_{3})   \ H_{d}\ S_{1}(r_{1}) S_{2}(r_{2}) \ S_{3}(r_{3})\ e^{-i\alpha \mathcal{J}_{3}}\ e^{-i\beta \mathcal{J}_{2}} e^{-i\gamma \mathcal{J}_{3}} \label{decomp}.
\end{eqnarray}
More analysis will be conducted to glean further information from these results and in accordance with the core components of the current system.

\section{Milburn dynamics and covariance matrix \label{sec 3}}
To study the intrinsic decoherence effect on our quantum system, we follow the Milburn Model (MM) \cite{Milb1991}. In MM, the evolution of the density $\rho(t)$ on a sufficiently small time scale is uncertain. As a result,  the system evolves from the state $\rho(t)$ to $\rho(t+\tau)$ via
\begin{eqnarray}
 \rho(t+\tau)=e^{-i\mathfrak{P}(\tau)H}\rho(t)e^{i\mathfrak{P}(\tau)H}
\end{eqnarray}
with the probability $p(\tau)$. It is worth emphasizing that if $ t = 0 $ and $\rho(0)$ is Gaussian, the state $\rho(t)$ is Gaussian for all $\tau$.  
This is  due to the  stochastic unitarity of Milburn evolution and the quadratic form of the Hamiltonian \cite{Milb1991,Milb}. In addition,   we have $\mathfrak{P}(\tau)\rightarrow \tau$ and $p(\tau)\rightarrow 1$ in standard quantum mechanics. Assuming a Poisson model for the stochastic dynamics of time steps, the state will be governed by the following Milburn master equation
\begin{equation}
\dot{\rho}(t)=\Gamma\left[ e^{-\frac{iH}{\Gamma}}\rho(0) e^{\frac{iH}{\Gamma}}-\rho(t)\right]
\end{equation}
where $\Gamma^{-1}$ is the decoherence rate defined as 
\begin{eqnarray}
 \Gamma^{-1}=\lim\limits_{\tau\rightarrow 0}\mathfrak{P(\tau)}.
\end{eqnarray}
The formal solution to Milburn's equation is given by \cite{Milb,urzua}
\begin{eqnarray}
\rho(t,\Gamma)&=&e^{-\Gamma t} \sum_{k=0}^{\infty}\frac{(\Gamma t)^k}{k!}e^{-\frac{ikH}{\Gamma}}\rho(0)e^{\frac{iHk}{\Gamma}}\label{milburn}.
\end{eqnarray}
Additionally, by using  (\ref{decomp}), one can show  the result
\begin{align}
	e^{-i\frac{k}{\Gamma}H}= e^{i\gamma \mathcal{J}_{3}}\ e^{i\beta \mathcal{J}_{2}}\ e^{i\alpha \mathcal{J}_{3}}\ S_{1}(-r_{1})\ S_{2}(-r_{2})\ S_{3}(-r_{3})   \ e^{-i\frac{k}{\Gamma}H_{d}}\ S_{1}(r_{1})\ S_{2}(r_{2})\ S_{3}(r_{3})\ e^{-i\alpha \mathcal{J}_{3}}\ e^{-i\beta \mathcal{J}_{2}}\ e^{-i\gamma \mathcal{J}_{3}}\label{diag}.
\end{align}
Since the coupling becomes ultra-strong, the ground state becomes populated with virtual excitations. This is due to the counter-rotating terms \cite{Rad2021oct,maintaining}, which simultaneously create the excitations in two modes. Then we will consider the ground state
$ \rho(0)=|000\rangle\langle000|$.
To go further, we collect the annihilation and creation operators of the three modes in the vector 
\begin{eqnarray}
\mathbb{A}=(\mathbb{A}_n)= (\hat{a}_1,\hat{a}_1^{\dagger},\hat{a}_2,\hat{a}_2^{\dagger},\hat{a}_3,\hat{a}_3^{\dagger})^{T}.
\end{eqnarray}
 The commutation relations  $[\hat{a}_i,\hat{a}^{\dagger}_j]=\delta_{ij}$, reduce to the following compact form
\begin{eqnarray}
 [\mathbb{A}_n,\mathbb{A}_m]=i\mathbb{J}_{n,m}, \quad i\mathbb{J}_{6\times6}=\bigoplus \tilde{\mathbb{I}}_{2\times 2}, \quad  \tilde{\mathbb{{I}}}_{2\times2}=\left(\begin{array}{cc}
     0&  1\\
     -1& 0
\end{array}\right).
\end{eqnarray}
Now, to put the transformations used above  in their symplectic form,  we recall that \cite{Redist}
\begin{eqnarray}
 \mathcal{U}\mathbb{A}\mathcal{U}^{-1}=\mathbb{S}\mathbb{A}\label{law}
\end{eqnarray}
 where the matrix $\mathbb{S}$ is called the symplectic representation of  $\mathcal{U}$ and verifies $\det(\mathbb{S})=1$ with $\mathbb{S}^{\dagger}\mathbb{J}\mathbb{S}=\mathbb{J}$. In our formalism, the covariance matrix reduces to
\begin{eqnarray}
\sigma_{n,m}&=&\langle \lbrace \mathbb{A}_{n},\mathbb{A}^{\dagger}_{m}
\rbrace\rangle_{\rho}-2\langle \mathbb{A}_{n}\rangle_{\rho}\langle\mathbb{A}^{\dagger}_{m}\rangle_{\rho}.
\end{eqnarray}
We mention that our state is centered, thus,  $\langle \mathbb{A}_{n}\rangle_{\rho}=\langle\mathbb{A}^{\dagger}_{m}\rangle_{\rho}=0$. On the other hand, we rewrite  (\ref{milburn}) in its symplectic form
\begin{eqnarray}
\sigma(t,\Gamma)= e^{-\Gamma t}\sum\limits_{k=0}^{\infty}\frac{(\Gamma t)^k}{k!}
 \mathbb{H} \sigma(0)\mathbb{H}^{\dagger}\label{haM}\label{product}
\end{eqnarray}
by showing that 
\begin{eqnarray}
 \sigma(0)=\text{diag}(1,1,1,1,1,1)
\end{eqnarray}
where {$(\cdot)^{\dagger}$ is the Hermitian conjugate of $(\cdot)$ and $\mathbb{H}$ is the symplectic representation of the operator $e^{-\frac{ikH}{\Gamma}}$. We use (\ref{diag}) to demonstrate that  the matrix $\mathbb{H}$ is
\begin{eqnarray}
 \mathbb{H}=\mathbb{S}_{12}(\gamma)\mathbb{S}_{13}(\beta)\mathbb{S}_{12}(\alpha)\mathbb{S}_{123}(-r_1,-r_2,-r_3)\mathbb{D}(\Gamma)\mathbb{S}_{123}(r_1,r_2,r_3)\mathbb{S}_{12}(-\alpha)\mathbb{S}_{13}(-\beta)\mathbb{S}_{12}(-\gamma)
\end{eqnarray}
where the symplectic matrices of $e^{iz \mathcal{J}_{3}}$ and $e^{i\beta \mathcal{J}_{2}}$ are given by
\begin{eqnarray}
\mathbb{S}_{12}(z)
&=&\begin{pmatrix}
   \cos(z) &  0&\frac{R+R^{-1}}{2}\sin(z)&\frac{R-R^{-1}}{2}\sin(z)&0&0\\
0  & \cos(z)&\frac{R-R^{-1}}{2}\sin(z)&\frac{R+R^{-1}}{2}\sin(z)&0&0\\
      -\frac{R+R^{-1}}{2}\sin(z)&\frac{R-R^{-1}}{2}\sin(z)&\cos(z)&0&0&0\\
   \frac{R-R^{-1}}{2}\sin(z)&-\frac{R+R^{-1}}{2}\sin(z)&0&\cos(z)&0&0\\
   0&0&0&0&1&0\\
   0&0&0&0&0&1
\end{pmatrix} \label{rot}
\\
  \mathbb{S}_{13}(\beta)
&=&\begin{pmatrix}
   \cos(\beta) & 0&0&0 &-\frac{\tilde{R}+\tilde{R}^{-1}}{2}\sin(\beta)&\frac{\tilde{R}-\tilde{R}^{-1}}{2}\sin(\beta)\\
0  & \cos(\beta)&0&0&\frac{\tilde{R}-\tilde{R}^{-1}}{2}\sin(\beta)&-\frac{\tilde{R}+\tilde{R}^{-1}}{2}\sin(\beta)\\
     0&0&1&0&0&0\\
      0&0&0&1&0&0\\
   \frac{\tilde{R}+\tilde{R}^{-1}}{2}\sin(\beta)&\frac{\tilde{R}-\tilde{R}^{-1}}{2}\sin(\beta)&0&0&\cos(\beta)&0\\
   \frac{\tilde{R}-\tilde{R}^{-1}}{2}\sin(\beta)& \frac{\tilde{R}+\tilde{R}^{-1}}{2}\sin(\beta)&0&0&1&\cos(\beta)\\
\end{pmatrix}
\end{eqnarray}
with $z=\alpha, \gamma$,  $R=\sqrt{\frac{\omega_1}{\omega_2}}$ and  $\tilde{R}=\sqrt{\frac{\omega_3}{\omega_1}}$ have been set. The symplectic representation of the three squeezers is reduced to 
\begin{eqnarray}
 \mathbb{S}_{123}(r_1,r_2,r_3)=\bigoplus\limits_{j=1}^{3}
\begin{pmatrix}
     \cosh(r_j)&\sinh(r_j)  \\
     \sinh(r_j)&\cosh(r_j)\\
\end{pmatrix}, \quad j=1,2,3\label{squeeez}.
\end{eqnarray}
Finally, the symplectic form of the transformation $e^{-i\frac{k}{\Gamma}H_{d}}$, which is a diagonal matrix $\mathbb{D}$ \cite{R1}, is obtained in a straightforward fashion. This is
\begin{eqnarray}
 \mathbb{D}(\Gamma)=\bigoplus\limits_{j=1}^{3}
\begin{pmatrix}
    e^{-ik\frac{\Omega_j}{\Gamma}}&0  \\
    0&  e^{ik\frac{\Omega_j}{\Gamma}}\\
\end{pmatrix}.
\end{eqnarray}
We mention that after performing the matrix product (\ref{product}), the infinite sum can be exactly computed by using the identity
\begin{eqnarray}
 e^{-\Gamma t}\sum\limits_{k=0}^{+\infty}\frac{(\Gamma t)^{k}}{k!}e^{if(\Omega_1,\Omega_2,\Omega_3)\frac{k}{\Gamma}}=e^{\Gamma t\left(e^{i\frac{f(\Omega_1,\Omega_2,\Omega_3)}{\Gamma}}-1\right)}
\end{eqnarray} 
where $f$ is a well defined function of normal frequencies $\Omega_j$.

\section{Quantum entanglement and  virtual excitations \label{sec 4}}

We analyze the dynamics and redistribution of virtual excitations between the three oscillators, together with the dynamics of bipartite entanglement. Both quantities are derived from the covariance matrix (\ref{product}), that is
\begin{eqnarray}
 \sigma(t,\Gamma)=\begin{pmatrix}
      \sigma_{a}&\sigma_{ab}& \sigma_{ac} \\
      \sigma_{ab}^{t}&\sigma_{b}&\sigma_{bc}\\
      \sigma_{ac}^{t}&\sigma_{bc}^{t}&\sigma_{c}
 \end{pmatrix}
\end{eqnarray}
where the single mode states are given by
\begin{eqnarray}
 \sigma_{a}=\begin{pmatrix}
      \sigma_{11}&\sigma_{12} \\
      \sigma_{12}^{\ast}&\sigma_{11}
  \end{pmatrix},\quad \sigma_{b}=\begin{pmatrix}
      \sigma_{33}&\sigma_{34} \\
      \sigma_{34}^{\ast}&\sigma_{33}
 \end{pmatrix}, \quad \sigma_{c}=\begin{pmatrix}
      \sigma_{55}&\sigma_{56} \\
      \sigma_{56}^{\ast}&\sigma_{55}
  \end{pmatrix}
\end{eqnarray}
and  the correlation matrices between two modes are
\begin{eqnarray}
 \sigma_{ab}=\begin{pmatrix}
      \sigma_{13}&\sigma_{14} \\
      \sigma_{23}^{\ast}&\sigma_{24}
  \end{pmatrix},\ \ \sigma_{ac}=\begin{pmatrix}
      \sigma_{15}&\sigma_{16} \\
      \sigma_{25}^{\ast}&\sigma_{26}
  \end{pmatrix}, \ \ \sigma_{bc}=\begin{pmatrix}
      \sigma_{35}&\sigma_{36} \\
      \sigma_{45}^{\ast}&\sigma_{46}
  \end{pmatrix}
\end{eqnarray}
where $t$ stands for transpose. By making use of the tracing-out prescription for Gaussian states in the covariance matrix formalism, we end up with the bipartite states \cite{adesso}
\begin{eqnarray}
 \sigma^{ab}=\begin{pmatrix}
      \sigma_{a}&\sigma_{ab} \\
      \sigma_{ab}^{t}&\sigma_{b}
 \end{pmatrix}, \quad 
\sigma^{ac}= \begin{pmatrix}
      \sigma_{a}&\sigma_{ac} \\
      \sigma_{ac}^{t}&\sigma_{c}
 \end{pmatrix}, \quad 
 \sigma^{bc}=\begin{pmatrix}
      \sigma_{b}&\sigma_{bc} \\
      \sigma_{bc}^{t}&\sigma_{c}
 \end{pmatrix} \label{bipartite}.
\end{eqnarray}
Due to the length of the explicit expressions for the matrix elements $\sigma_{nm}$ ($n,m=1,\cdots, 6$), we only report their formal relations and numerical results.

We investigate the impact of virtual excitations on quantum entanglement in a harmonic oscillator system with ultra-strongly coupled. We show that the average number of excitations in the ground state is
\begin{eqnarray}
\langle N_{j}\rangle&=&\langle \hat{a}_{j}^{+}\hat{a}_{j}\rangle=\frac{1}{2}(\sigma_{2j-1,2j-1}-1), \quad  j=1,2,3.
\end{eqnarray}
Additionally, to quantify the bipartite entanglement $(k|l)$ while $k,l=a,b,c$. The bipartite covariance matrix  $\sigma^{kl}$ (\ref{bipartite})  are  partially transposed   to obtain  $\tilde{\sigma}$. The positive minimal symplectic   eigenvalue is given by \cite{adesso1}
\begin{eqnarray}
\tilde{\nu}_{min}^{kl}=\sqrt{ \frac{1}{2}\left(\Delta_{kl}-\sqrt{\Delta_{kl}^{2}-4\det \sigma^{kl}}\right)}, \quad k,l=a,b,c
\end{eqnarray}
where  the seralian $\Delta_{kl}$ is defined as 
\begin{eqnarray}
\Delta_{kl}=\det\sigma_{k}+\det\sigma_{l}-2\det\sigma_{kl}.
\end{eqnarray}
Consequently, the logarithmic negativity reduces to
\begin{eqnarray}
E_{kl}=\max\left(0,-\ln(\tilde{\nu}^{kl}_{min})\right).
\end{eqnarray}
In contrast to non-classical correlations that go beyond inseparability, entanglement is monogamous among the parties of the system.  This entails that entanglement is not freely shared between the parties of the whole system. In particular, for a tripartite system, the monogamy of entanglement was formulated in  \cite{CKW1}
and shown in \cite{ CKW2}, that is 
\begin{eqnarray}
 E_{k|lm} \geq  E_{kl}+E_{km}, \quad k,l,m\in \lbrace a,b,c \rbrace
\end{eqnarray}
where $E_{kl} $ represents the bipartite entanglement of $k$ and $l$, the vertical bar denotes the bipartite split. Motivated by the results of \cite{maintaining,Rad2021oct}, which demonstrate the interaction between entanglement and virtual excitations, we ask a fundamental question: whether virtual excitations are distributed in the same way that entanglement is. 


\section{Numerical results and discussions\label{sec 5}}
\subsection{Isotropic case}

Assume you have a resonant system with $\omega_1=\omega_2=\omega_3=\omega_r$ and $J_{12}=J_{13}=J_{23}=J$. As a result, the standard frequencies become
\begin{align}
	\Omega_1=\sqrt{\omega_r^2+2J}, \quad
	\Omega_2=\Omega_3=\sqrt{\omega_r^2-J}.
\end{align} 
The matrix form of the rotation operator that leads to the diagonalized Hamiltonian is 
\begin{eqnarray}
 \mathcal{R}&=&\left(\begin{array}{ccc}
     \frac{1}{\sqrt{3}} &  \frac{1}{\sqrt{3}} & \frac{1}{\sqrt{3}} \\
      0&  \frac{1}{\sqrt{2}}& -\frac{1}{\sqrt{3}}\\
      -\sqrt{\frac{2}{3}}& \frac{1}{\sqrt{6}}& \frac{1}{\sqrt{6}}
 \end{array}\right)
\end{eqnarray}
and the rotation angles are reduced to
\begin{eqnarray}
\gamma&\to&-\arctan\left(\frac{1}{3}\right)\\
 \alpha&\to&\arctan\left(\sqrt{\frac{3}{2}}\right) \\
  \beta&\to& 
 \arccos\left( \frac{1}{\sqrt{6}}\right).
\end{eqnarray}
Then, use the fact that $2\sinh(r)\cosh(r)=\sinh(2r)$, we show the following virtual excitations 
\begin{eqnarray}
\langle N_1\rangle(t)&=& \frac{24}{50}{\cosh^2(r_1)}{ \sinh^2(r_1)}\left[1-\exp[-\Gamma t(1-\cos(2\Omega_1/\Gamma))]\cos(\Gamma t\sin(2\Omega_1/\Gamma))\right]\\\notag
&+&\frac{76}{50}{\cosh^2(r_2)}{ \sinh^2(r_2)}\left[1-\exp[-\Gamma t(1-\cos(2\Omega_2/\Gamma))]\cos(\Gamma t\sin(2\Omega_2/\Gamma))\right]\\
\langle N_2\rangle(t)&=& \frac{1}{50}{\cosh^2(r_1)}{
	\sinh^2(r_1)}\left[1-\exp[-\Gamma t(1-\cos(2\Omega_1/\Gamma))]\cos(\Gamma t\sin(2\Omega_1/\Gamma))\right]\\\notag\\\notag
&+&\frac{99}{50}{\cosh^2(r_2)}{\sinh^2(r_2)}\left[1-\exp[-\Gamma t(1-\cos(2\Omega_2/\Gamma))]\cos(\Gamma t\sin(2\Omega_2/\Gamma))\right]\\
\langle N_3\rangle(t)&=& \frac{3}{2}{\cosh^2(r_1)}{\sinh^2(r_1)}\left[1-\exp[-\Gamma t(1-\cos(2\Omega_1/\Gamma))]\cos(\Gamma t\sin(2\Omega_1/\Gamma))\right]\\\notag\\\notag
&+&\frac{1}{2}{\cosh^2(r_2)}{\sinh^2(r_2)}\left[1-\exp[-\Gamma t(1-\cos(2\Omega_2/\Gamma))]\cos(\Gamma t\sin(2\Omega_2/\Gamma))\right].
\end{eqnarray}
Additionally, by virtue of the above expressions, we show the steady values of excitations, such that 
\begin{eqnarray}
 \langle N_1\rangle_{\infty}&=& \frac{24}{50}{\cosh^2(r_1)}{\sinh^2(r_1)}+\frac{76}{50}{\cosh^2(r_2)}{\sinh^2(r_2)}=\frac{J^2(43\omega_r^2+14J)}{200\omega_r^2(\omega_r^2+2J)(\omega_r^2-J)}\\\notag
 \langle N_2\rangle_{\infty}&=& \frac{1}{50}{\cosh^2(r_1)}{ \sinh^2(r_1)}+\frac{99}{50}{\cosh^2(r_2)}{\sinh^2(r_2)}=\frac{J^2(103\omega_r^2+149J)}{800\omega_r^2(\omega_r^2+2J)(\omega_r^2-J)}\\\notag
 \langle N_3\rangle_{\infty}&=& \frac{3}{2}{\cosh^2(r_1)}{\sinh^2(r_1)}+\frac{1}{2}{\cosh^2(r_2)}{ \sinh^2(r_2)}=\frac{J^2(13\omega_r^2-10J)}{32\omega_r^2(\omega_r^2+2J)(\omega_r^2-J)}
\end{eqnarray}
which are independent of $\Gamma$. Note also that the three steady values will be equal when the coupling takes the value $J=\omega_r^2/2$. Furthermore, it is simple to demonstrate that $\langle N_j\rangle $ are increasing functions with respect to $J$. Furthermore, we show $\langle N_3\rangle_{\infty}>\langle N_1\rangle_{\infty}>\langle N_2\rangle_{\infty}$ for all $J<\omega_r^2/2$, and the hierarchy will be inverted, i.e., $\langle N_3\rangle_{\infty}<\langle N_1\rangle_{\infty}<\langle N_2\rangle_{\infty}$ for $J>\omega_r^2/2$. We also mention that the dynamics reaches a steady state during times of order
\begin{eqnarray}
 t_{steady}\sim\max\left[\frac{1}{\Gamma(1-\cos(2\Omega_1/\Gamma))},\frac{1}{\Gamma(1-\cos(2\Omega_2/\Gamma))}\right].
\end{eqnarray}

As shown in Fig. \ref{figB}, as the coupling increases, $t_{steady}$ increases. Additionally, we observe that excitations increase as the coupling becomes ultra-strong. Furthermore, as previously discussed, the steady values of excitations will be equal as long as $J=\omega_r^2/2$ and will become hierarchical if $J\neq\omega_r^2/2$. 


\begin{figure}[H]
	\centering	
	\includegraphics[width=8cm, height=5cm]{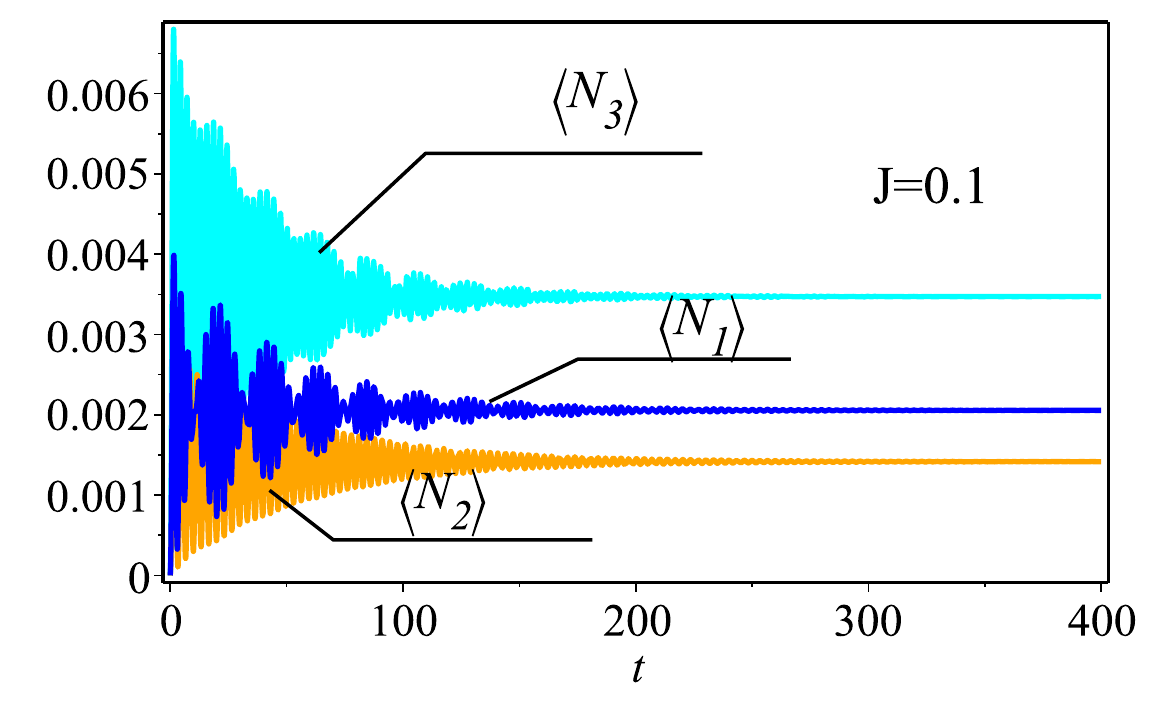}
	\includegraphics[width=8cm, height=5cm]{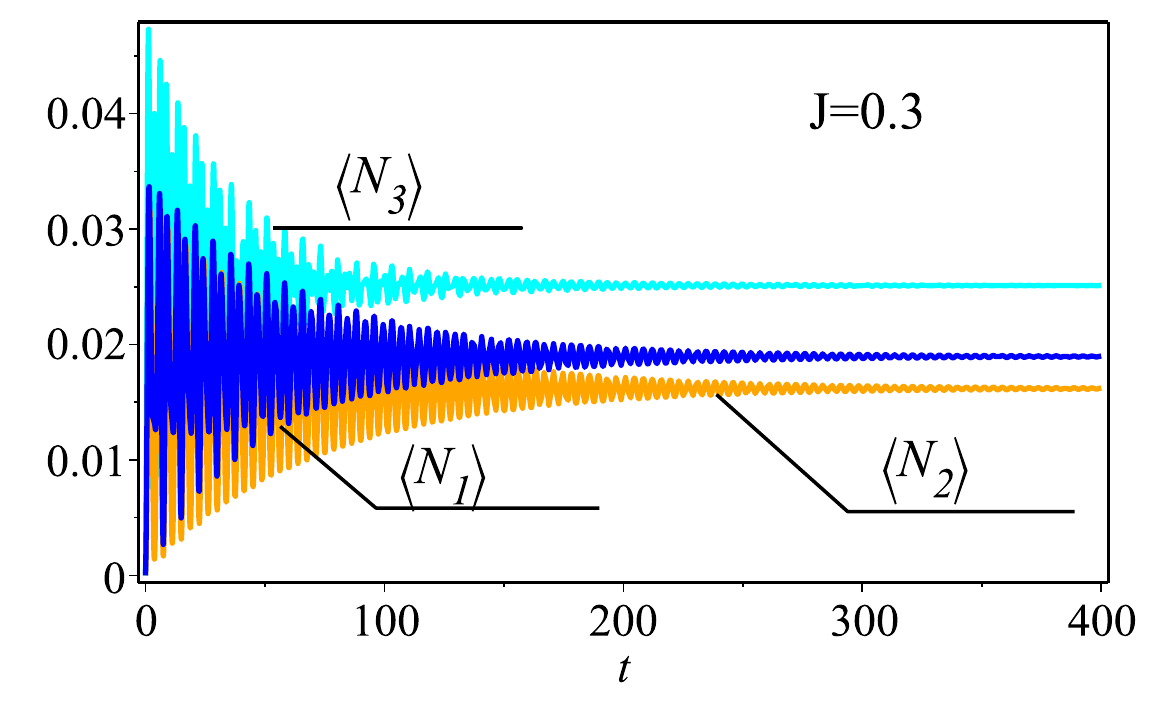}
	\includegraphics[width=8cm, height=5cm]{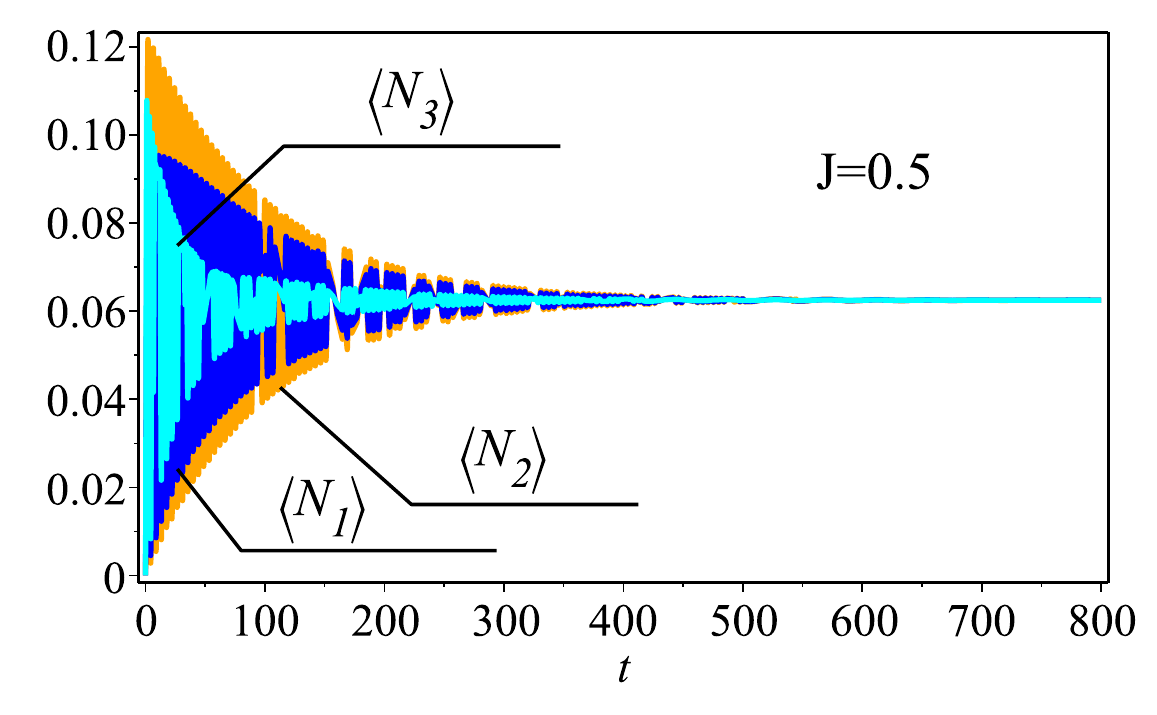}
\includegraphics[width=8cm, height=5cm]{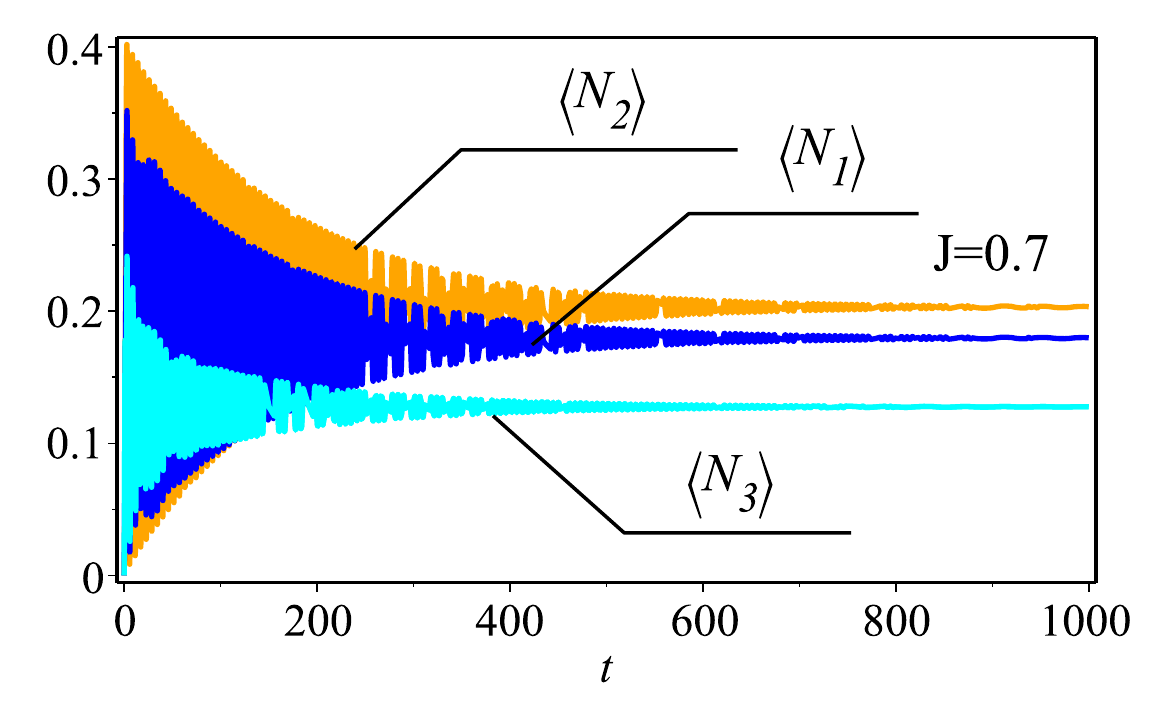}
		\caption{(color online) The effects of  coupling $J$ on the time of the transient state. The parameters $\omega_r=1$ and $\Gamma=100$ were set. }\label{figB}
\end{figure}

\subsection{Effects of Milburn decoherence}

To show the effects of the intrinsic decoherence on the dynamics of excitations, we plot in Fig. \ref{fig1} the dynamics  of the average of excitations $\langle N_{k} \rangle=\langle\hat{a}_{k}^{+}\hat{a}_{k}\rangle $. We set $\omega_r=\omega_1=\omega_2=\omega_3=1$,  $J_{13}=J_{12}=0.1$ and $J_{23}=0$ (indicating an open chain). The three populations simultaneously appear and undergo damped oscillatory behaviour to a steady state of excitation. This multi-oscillatory regime is due to normal frequencies $\Omega_k (k=1,2,3)$ and $\Gamma$. Besides, the three plots show that the excitations in particles $ a $ and $ b $ are the same  which implies that the excitations exchange between them.
We see that the excitations satisfy the inequalities $\langle N_{3} \rangle\leq \langle N_{1,2} \rangle$ all the time. Furthermore, we notice that as long as  $\Gamma$ is large, the stronger the excitations are. The Schr\"{o}dinger dynamics is obtained for $\Gamma\rightarrow\infty$, where the excitations undergo an undamped oscillatory behavior, and the steady excitations disappear. We also notice that as $\Gamma$ decreases, the oscillations disappear and the excitations exponentially revive.

\begin{figure}[H]
	\centering	
	\includegraphics[width=8.5cm, height=5.5cm]{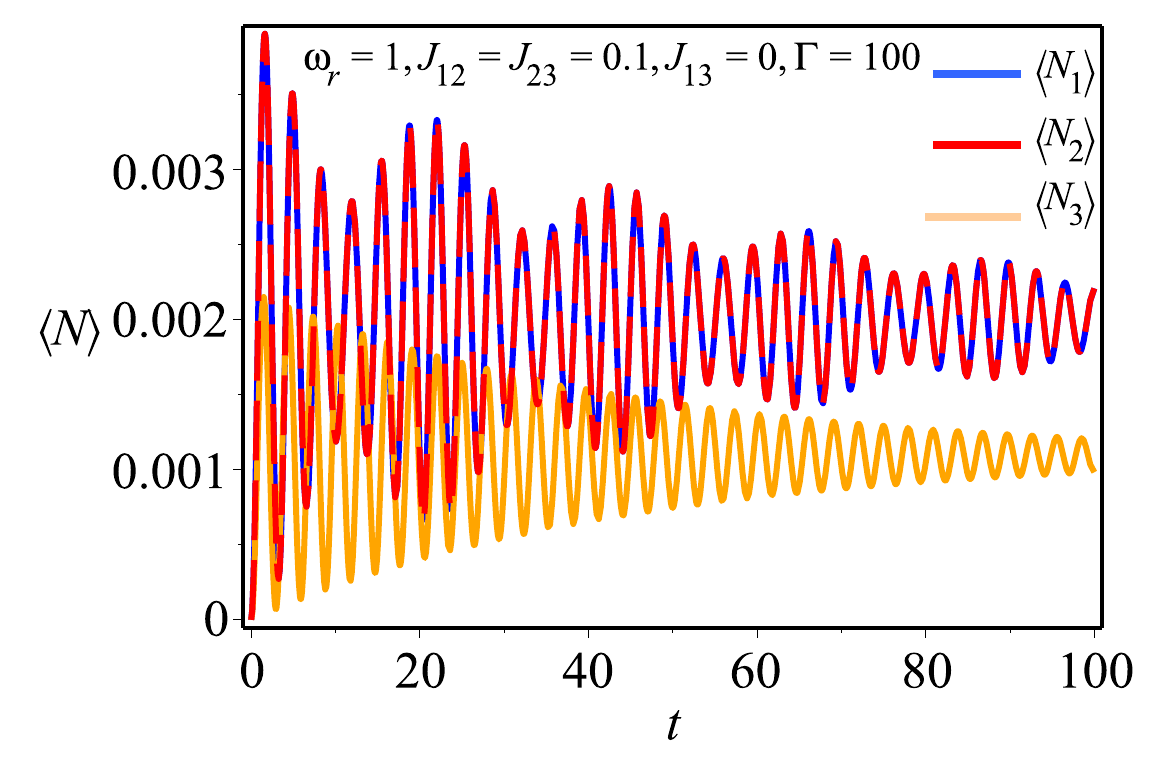}
	\includegraphics[width=8.5cm, height=5.5cm]{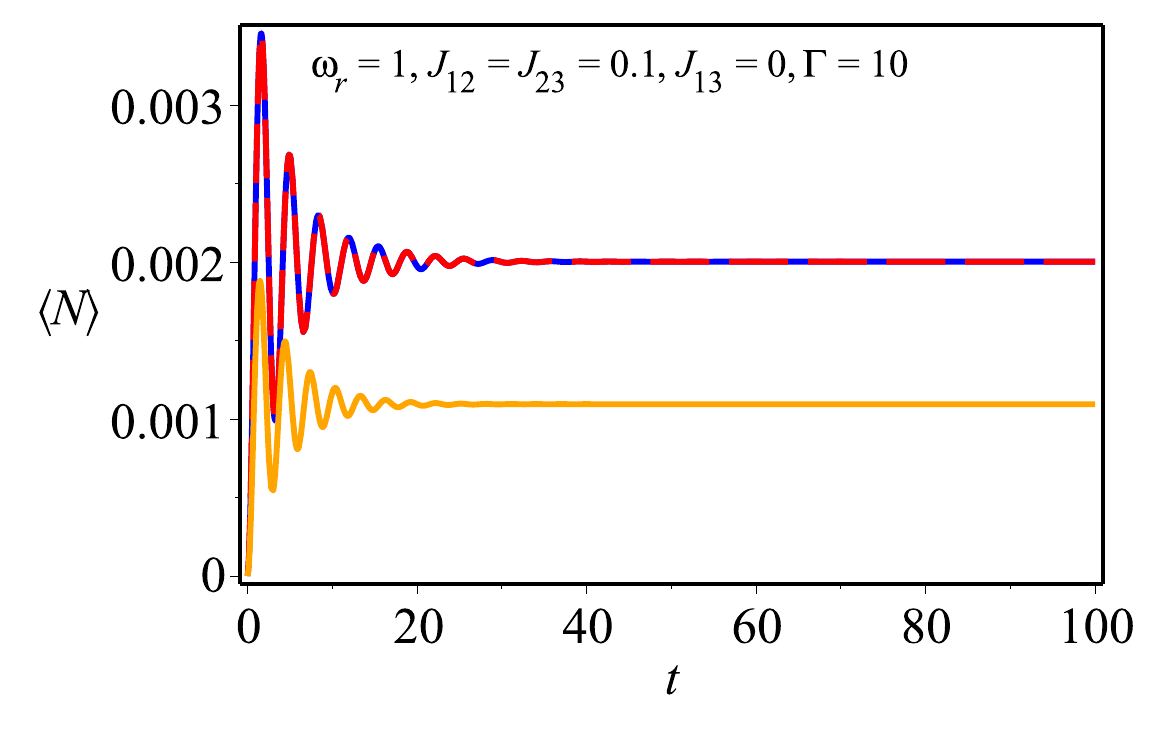}
	\includegraphics[width=8.5cm, height=5.5cm]{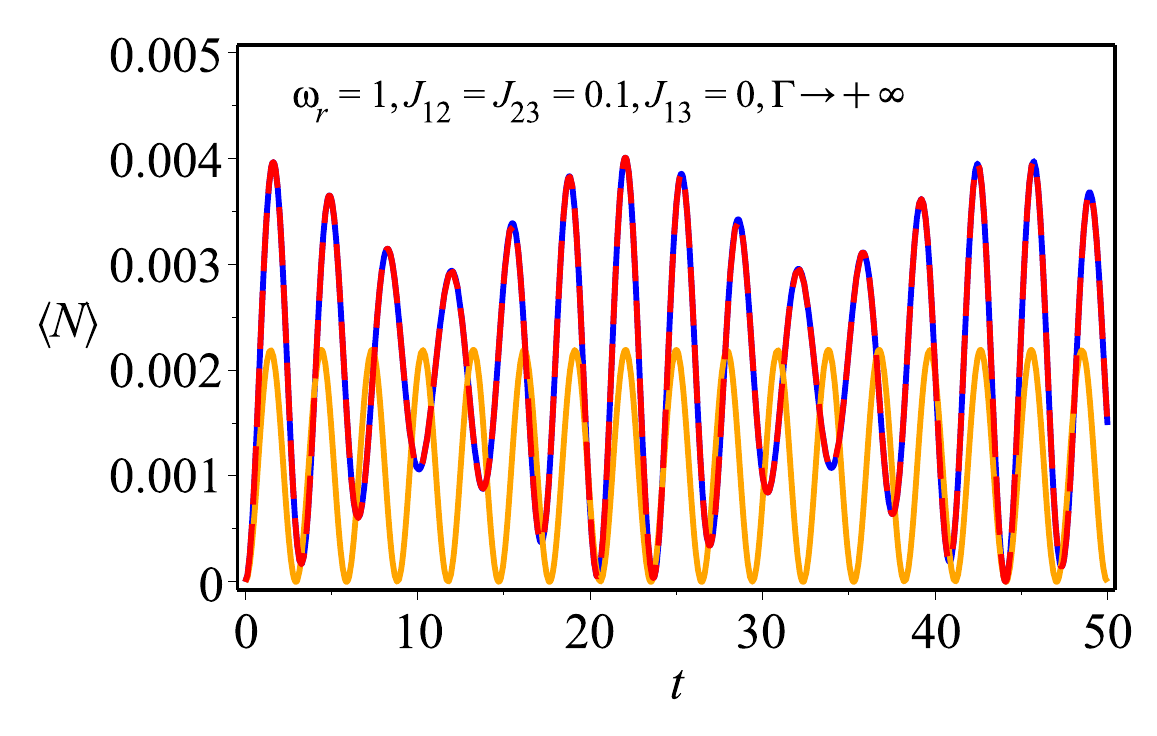}
		\includegraphics[width=8.5cm, height=5.5cm]{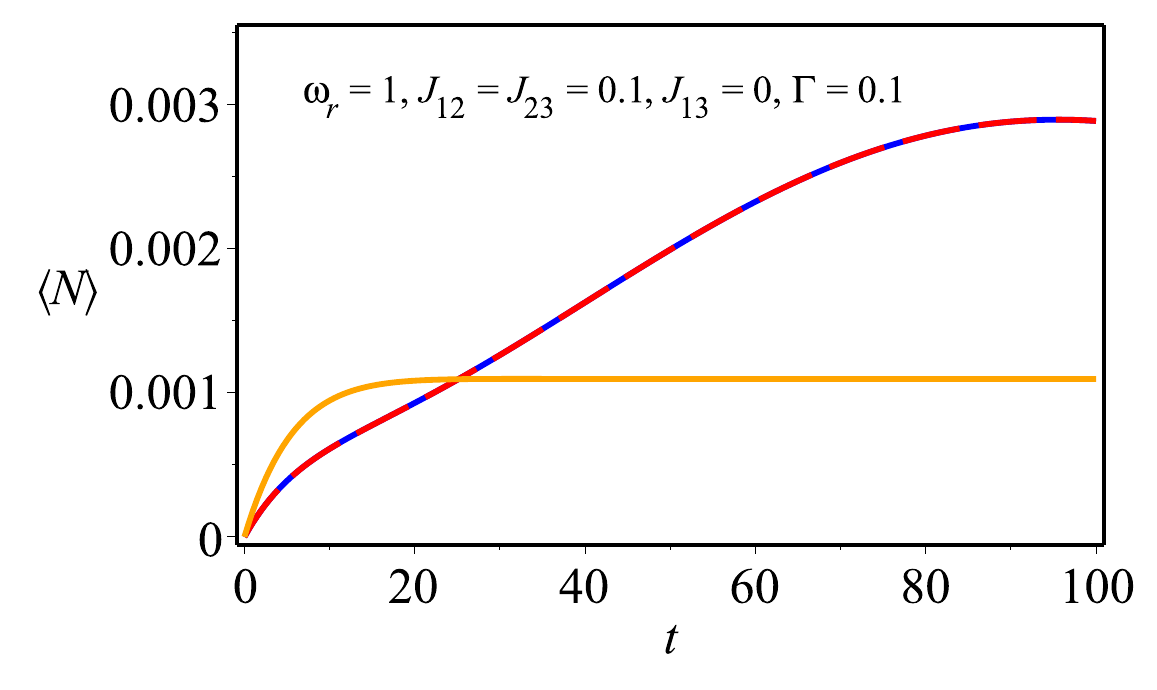}
		\caption{(color online) The effect of decoherence rate $\Gamma$ on  the dynamics of virtual excitations. The parameters $\omega_1=\omega_2=\omega_3=\omega_r=1$, $J_{23}=J_{12}=0.1$, and  $J_{13}=0$ were set. The case where $\Gamma\to +\infty$, corresponds to Schr\"{o}dinger dynamics. }\label{fig1}
\end{figure}

\subsection{Effects of coupling $J_{13}$ and central frequency $\omega_2$ }

To investigate the evolution
	of excitations as a function of the coupling $J_{13}$ and the central frequency $\omega_2$, we plot the dynamics of excitations in three particles by varying $J_{13}$ and $\omega_2$, respectively, in Figs. \ref{fig2} and \ref{fig3}.
First,  we set $\omega_r=\omega_1=\omega_2=\omega_3=1$,  $J_{13}=J_{12}=0.1$ as well as varying the coupling between particles $a $ and $c $. As expected,  because of the coupling $J_{13}$,   the excitations in particle $ c $ become more intensive than those in particles $ a $ and $ b $. 
Furthermore, the higher the {coupling  $J_{13}$, the more excitations are significant in particles $ a $ and $ c $, implying that excitations are transferred between them. 
Now, we weakly couple the particles $(a|b)$ and $(b|c)$,  and ultra-strongly couple the particles $(a|c)$. The frequencies of particles a and c are assumed to be equal to $\omega_r=1$, and we vary the central frequency $\omega_2$. As we move away from resonance, the excitations in particle c become more intense, owing to $J_{13}$. 
 Additionally, we notice an inverted dynamics for $\langle N_1\rangle $ and $\langle N_2\rangle $, which means the transfer of excitations between them. Decreasing the frequency $\omega_2$ to $0.8$, we observe the decrease of excitations in all particles. Fulfilling the condition $\omega_1+\omega_3=4\omega_2$, we observe a resonance, which amplifies the excitations in particles $ a $ and $ b $, and diminishes them in particle $ c $. 
 It is also worth noting that by lowering the central frequency to $\omega_2=0.3$, we observe an amazing extinction of excitations in particle $ b $ as well as a transfer of excitations between particles $ a $ and $ c $. These results show that, when excitations are important in a particle, they will be less important in at least one of the remaining particles. This indicates that excitations are not freely generated or annihilated. For this, we elucidate the polygamy of excitations among the three parties of the whole system.

\begin{figure}
	\centering	\includegraphics[width=8cm, height=5cm]{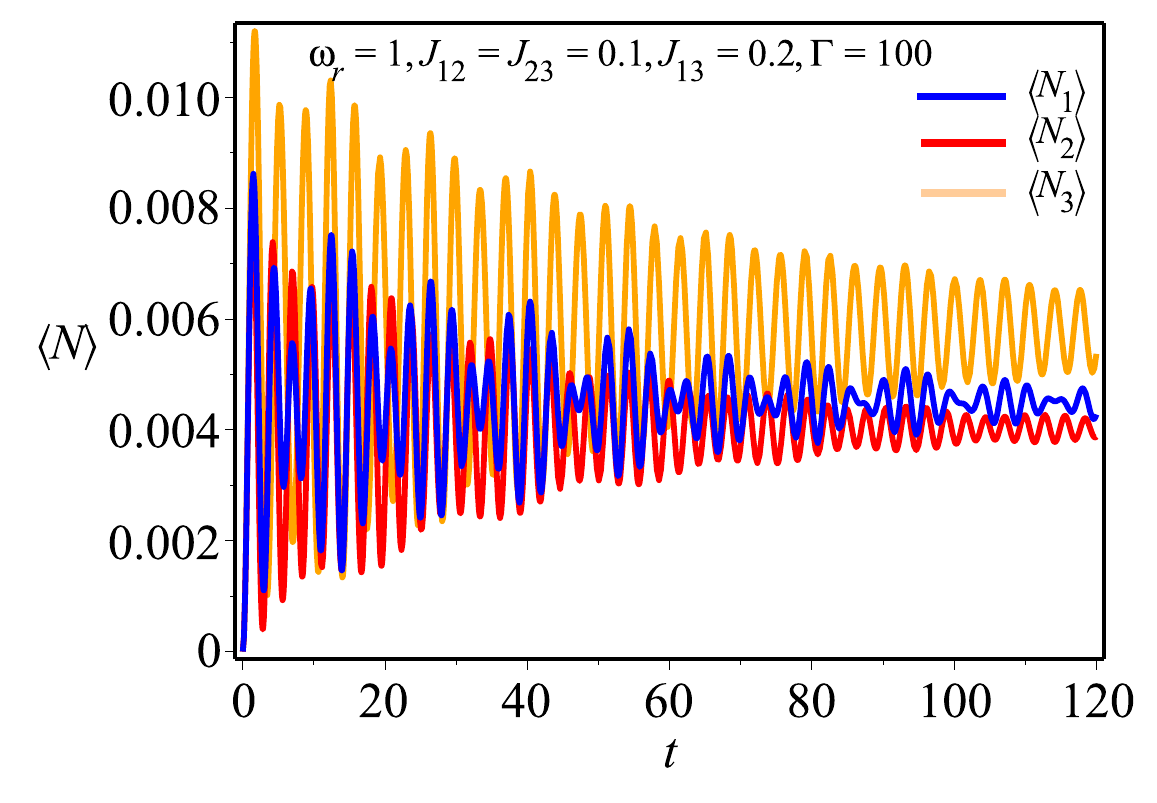}
	\includegraphics[width=8cm, height=5cm]{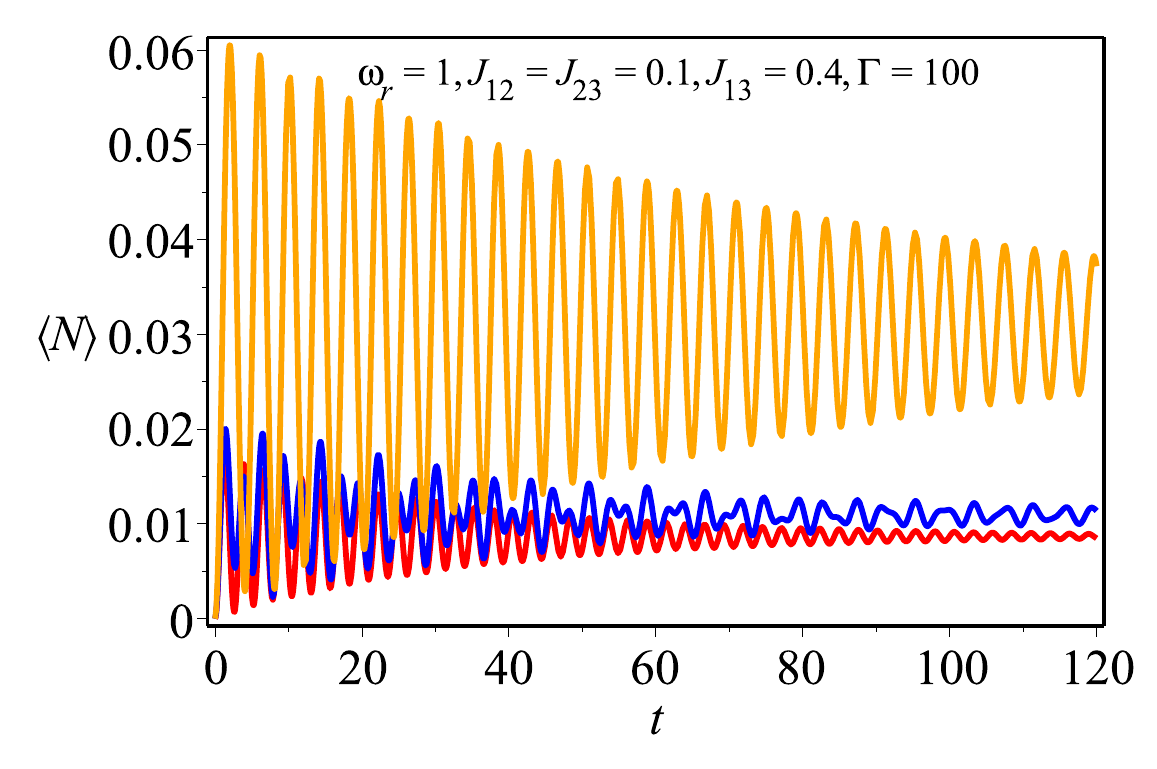}
	\includegraphics[width=8cm, height=5cm]{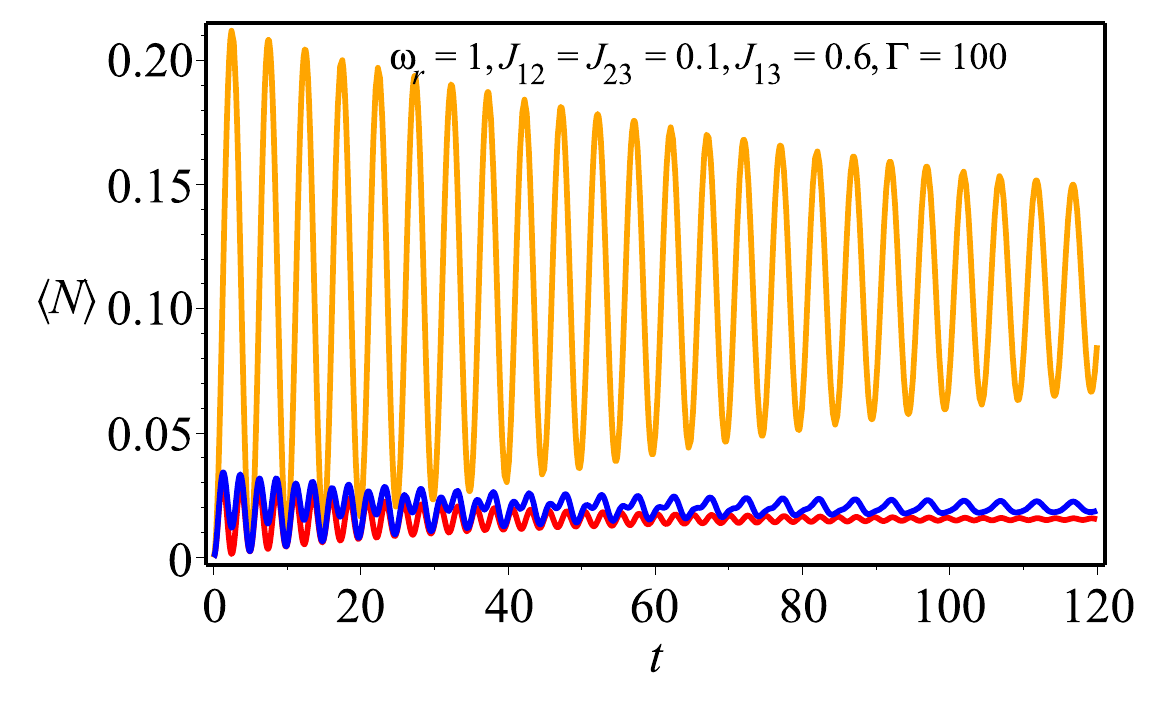}
		\includegraphics[width=8cm, height=5cm]{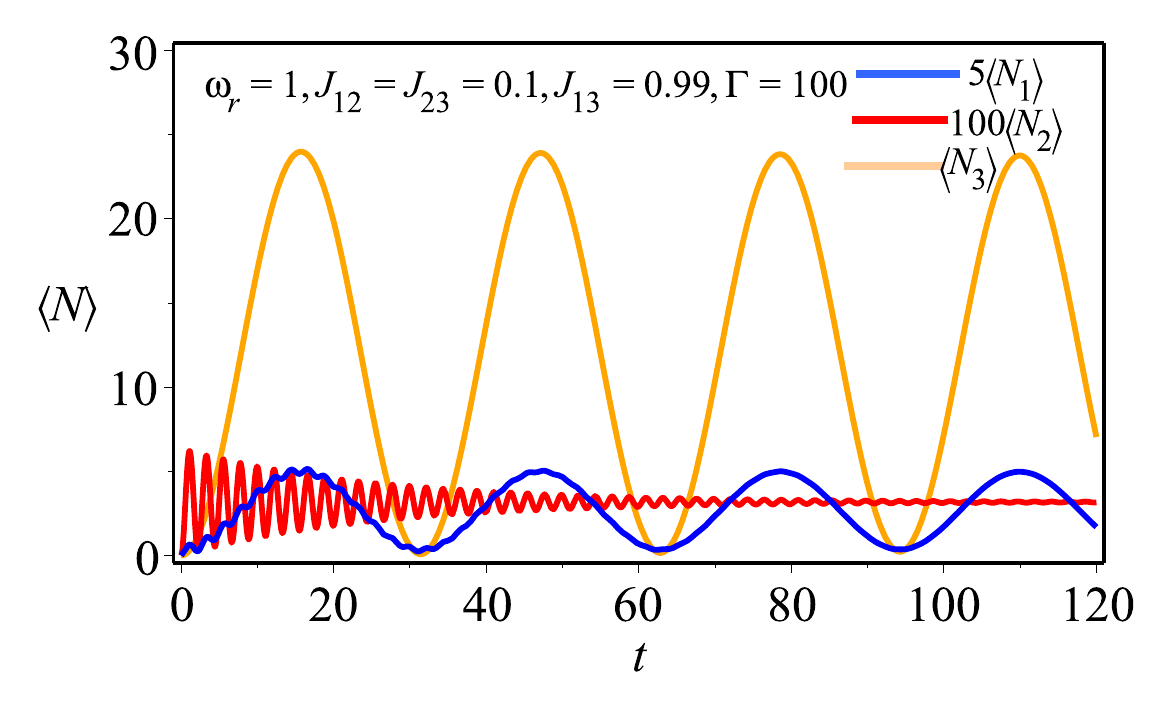}
		\caption{(color online) The effect of central coupling $J_{13}$ on  dynamics of virtual excitations. The parameters $\omega_1=\omega_2=\omega_3=\omega_r=1$,$J_{23}= J_{12}=0.1$ and $\Gamma=100$ were set. }\label{fig2}
\end{figure}
\begin{figure}
	\centering	\includegraphics[width=8cm, height=5cm]{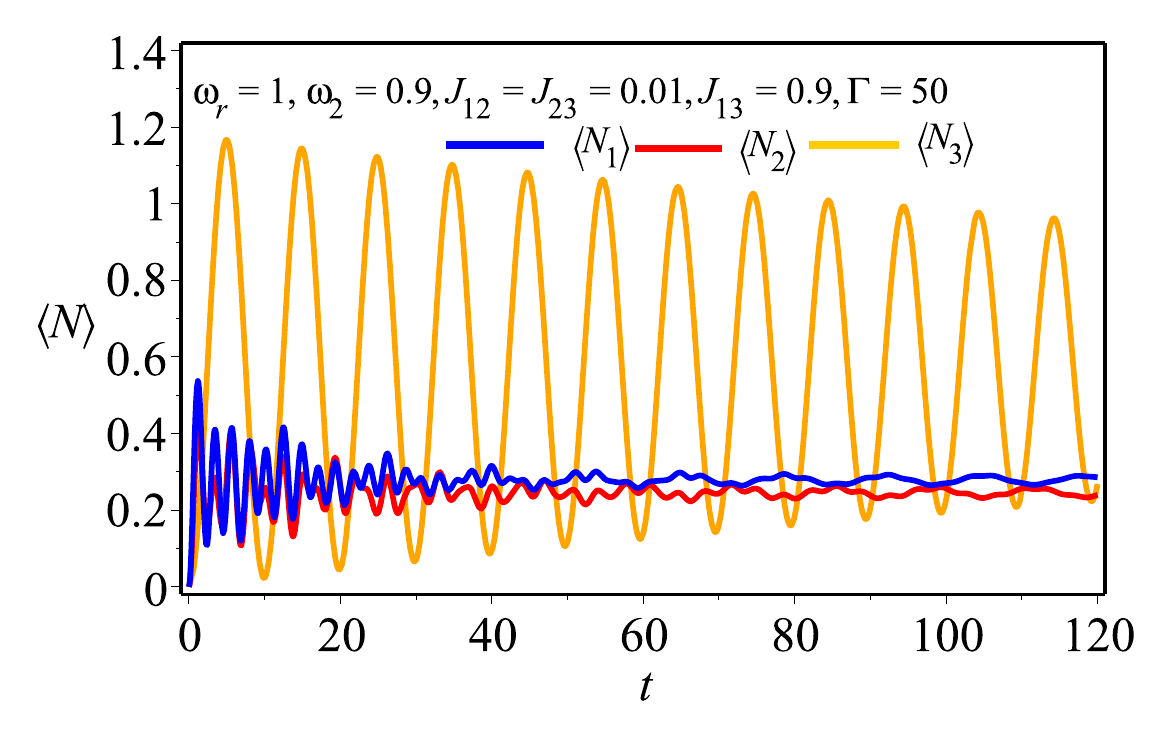}
	\includegraphics[width=8cm, height=5cm]{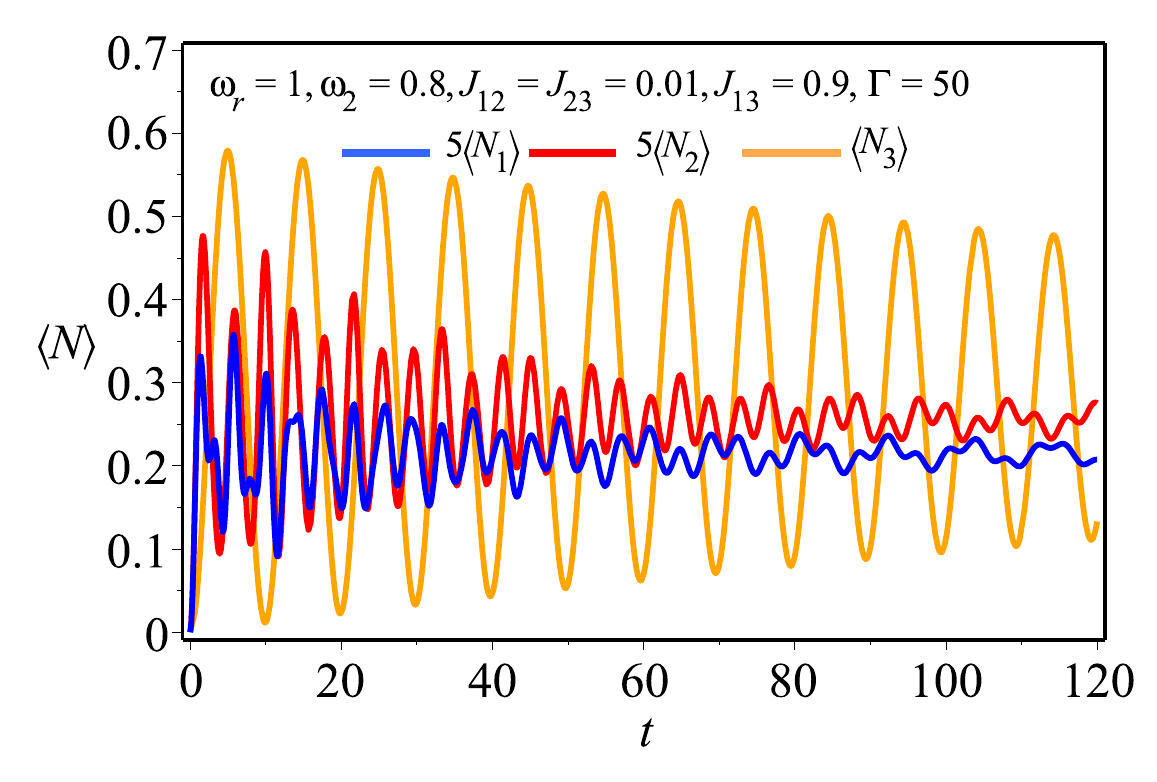}
	\includegraphics[width=8cm, height=5cm]{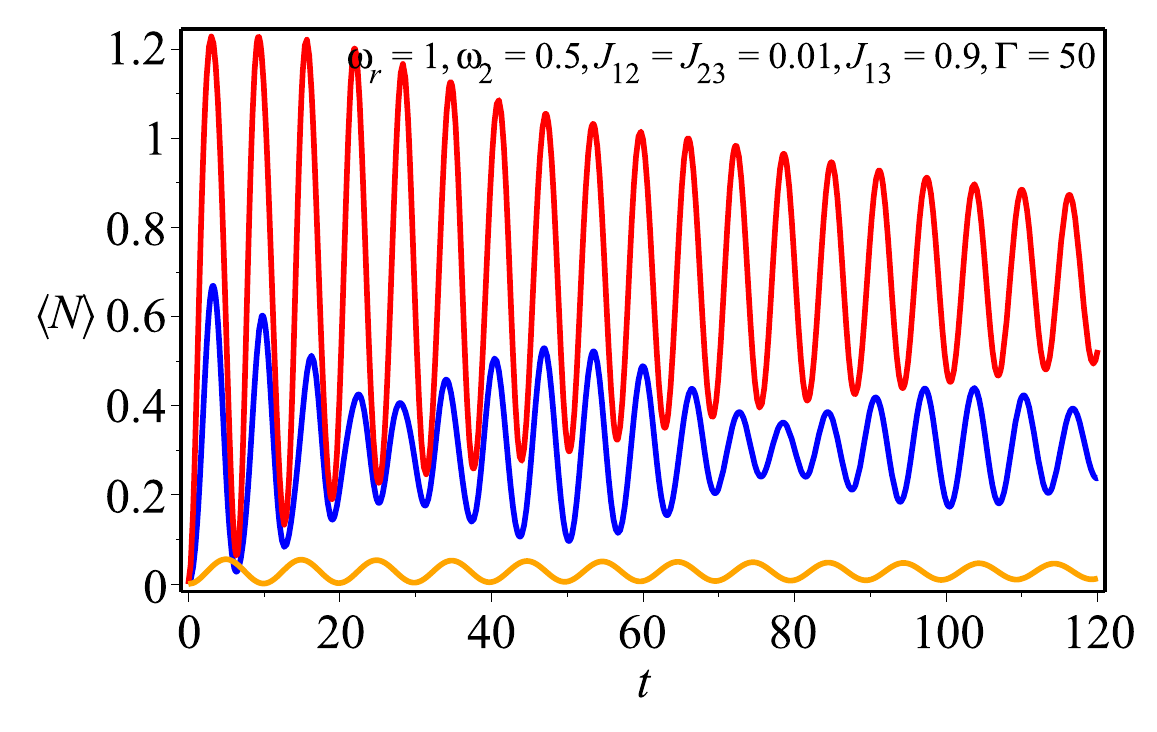}
		\includegraphics[width=8cm, height=5cm]{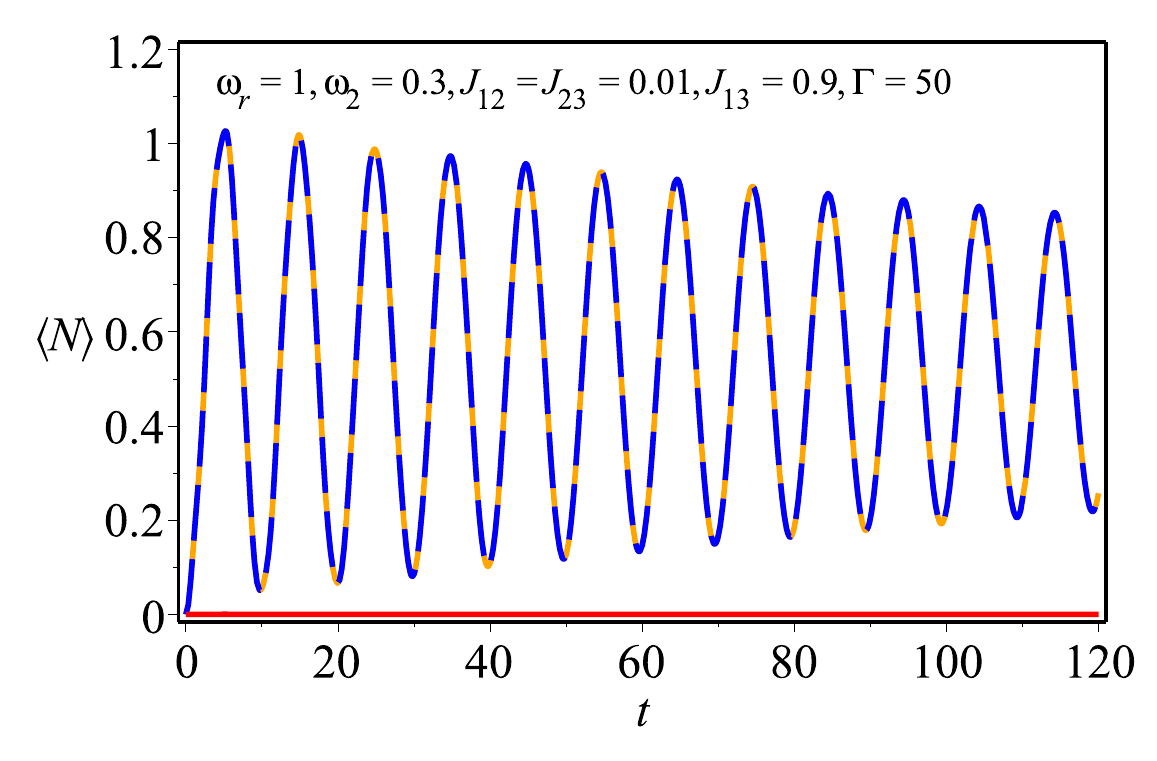}
		\caption{(color online) The effect of central frequency $\omega_2$ on the dynamics of virtual excitations. The parameters $\omega_1=\omega_3=\omega_r=1$,$J_{23}= J_{12}=0.01$, $J_{13}=0.9 $ and $\Gamma=50$ were set. }\label{fig3}
\end{figure}

\subsection{Virtual excitations and polygamy}

To investigate the interplay between excitations in  the bi-partitions of three oscillators. We define the bipartite virtual excitations in modes $k$ and $l$ as follows
\begin{eqnarray}
 N_{k|l}&=& \sqrt{\langle N_{k} \rangle \langle N_{l} \rangle}.
 \end{eqnarray}
 This can be seen as a  geometric mean of the number of excitations. As a result, the excitations in the bi-partition vanish if and only if they vanish at least in one mode. The present choice will be justified later, when the interconnection between excitations and entanglement is addressed. Similarly, we define the excitation in the partition $(k|lm)$ by
 \begin{eqnarray}
  N_{k|lm}&=& \sqrt{\langle N_{k}  \rangle N_{l|m} }.
 \end{eqnarray}
The excitations will be polygamous \cite{polygamy} if they satisfy the following  triangular inequality
\begin{eqnarray}
 N_{k|lm}\leq N_{k|l}+N_{k|m}\label{polygamy}.
\end{eqnarray}
These inequalities constrain the generation and extinction of excitations. This finding can be used to consider virtual excitations as a kind of quantum correlation \cite{enta1, Rad2021oct,maintaining}. Furthermore, the virtual excitations can be used to quantify quantum resources beyond entanglement. Here, to show the interplay between excitations, we plot in Fig. \ref{fig5}, the dynamics of the trade-off quantities
\begin{eqnarray}
 \delta^{i}_{jk}&=& N_{i|j}+N_{i|k}-N_{i|jk}, \quad k,l,m=a,b,c.
\end{eqnarray}

We numerically show that excitations are polygamous. This entails that excitations are not freely generated or annihilated between the several parties of the whole system. However, they undergo the triangular inequality given in (\ref{polygamy}).

\begin{figure}[H]
	\centering	\includegraphics[width=8cm, height=5.2cm]{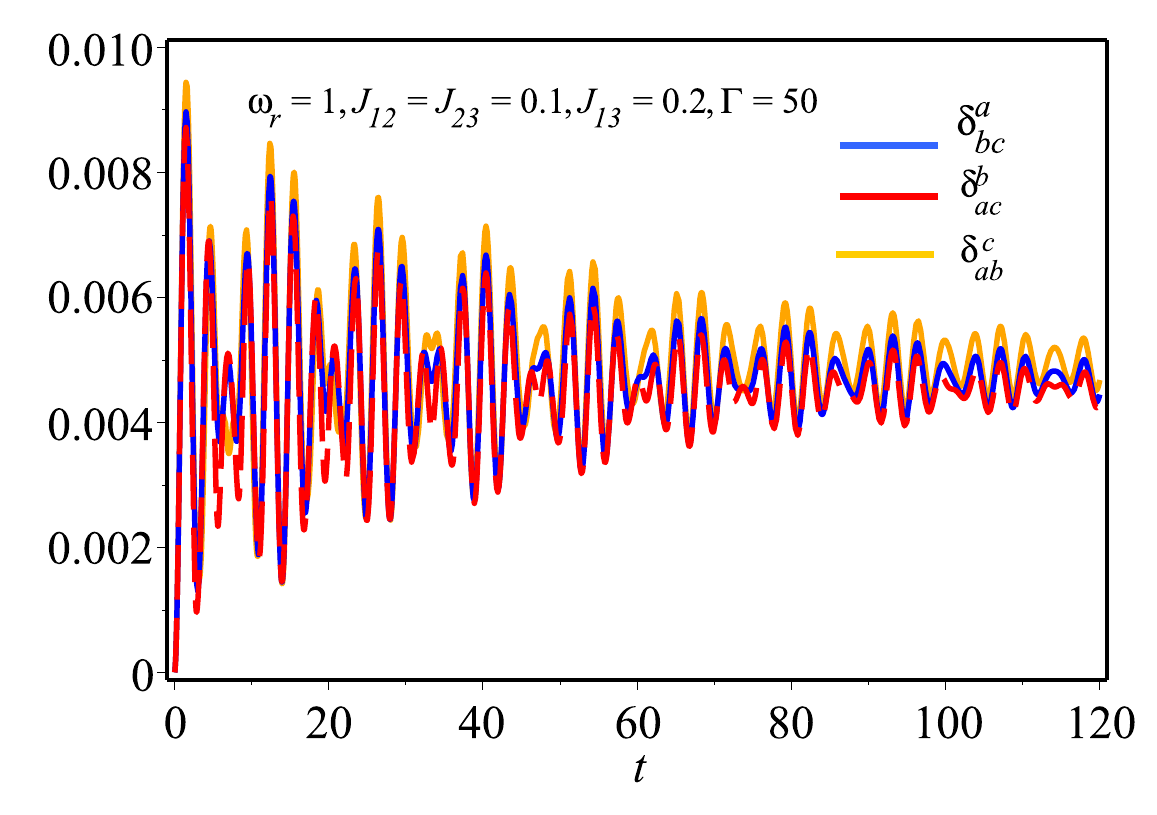}
		\centering	\includegraphics[width=8cm, height=5cm]{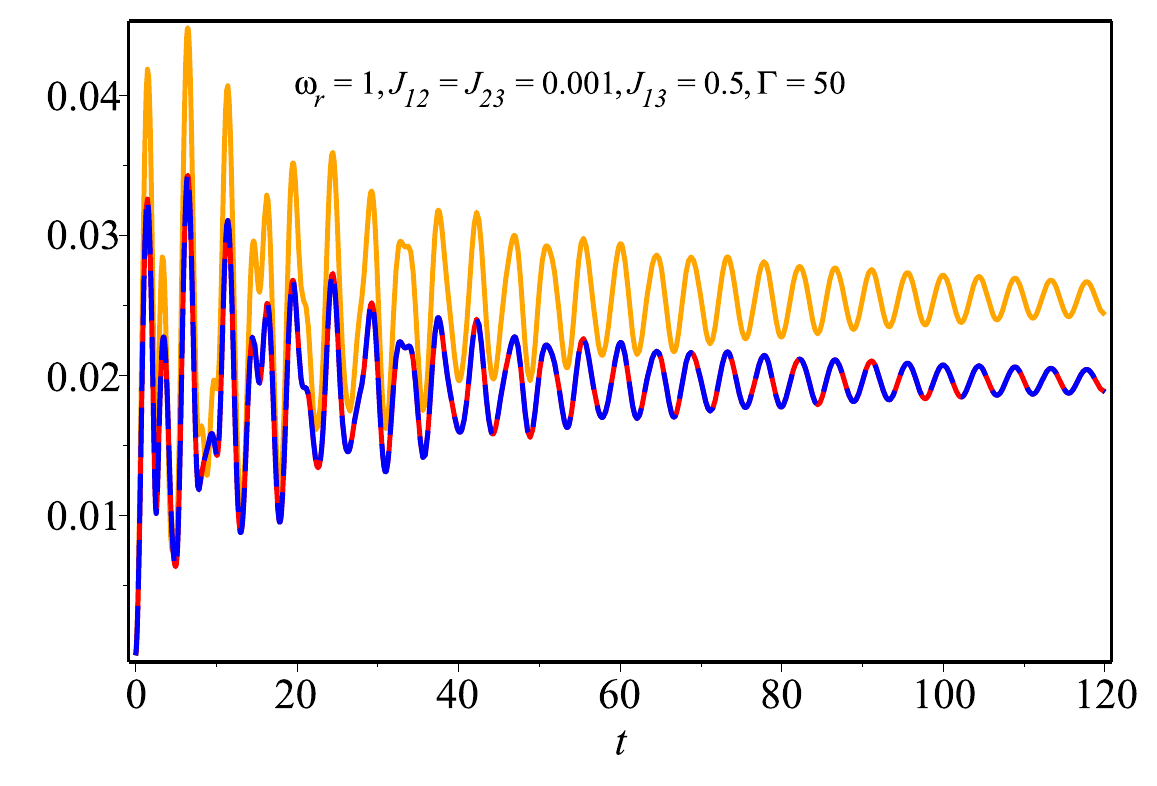}
		\caption{(color online) The illustration of polygamy of excitations. The parameters $\omega_1=\omega_2=\omega_3=\omega_r=1$, and  $\Gamma=50$ were set. }\label{fig5}
\end{figure}

\subsection{Entanglement and virtual excitations}

We tackle in this paragraph the interconnection between excitations and entanglement. For this aim, we plot in  Fig. \ref{fig6}  the dynamics of entanglement and excitations.
First, we assume an open chain $J_ {13} =0$ and ultra-strongly coupled oscillators $J_ {12} =J_ {23} =0.1$. Surprisingly, the bi-partition $a|b$ is more entangled than $b|c$. This originates from the number of excitations in those bi-partitions, i.e., $N_{a|b} \leq N_{b|c}$. Another result that deserves attention is the entanglement in the bi-partition $(a|c)$. The oscillators are indeed decoupled at $J_{13}=0$, but they are indirectly coupled via the oscillator $b$. As a result, the numbers of excitations in the partitions $a|c$ and $b|c$ are equal, and  $ E_{ac}=E_{bc}$. Second, we set the oscillators $a$ and $b$ to $J_{13}=J_{23}=0$ and strongly coupled them. Thus, the oscillator $c$ does not contain excitations. As a result, the partitions $j|c$, while  $(j\neq c)$ are not entangled and do not contain excitations. Finally, it is worthwhile to mention that both profiles exhibit similar dynamics. This points out that excitations can be used as quantifiers of entanglement \cite{Rad2021oct}.

\begin{figure}[H]
	\centering	\includegraphics[width=8cm, height=5cm]{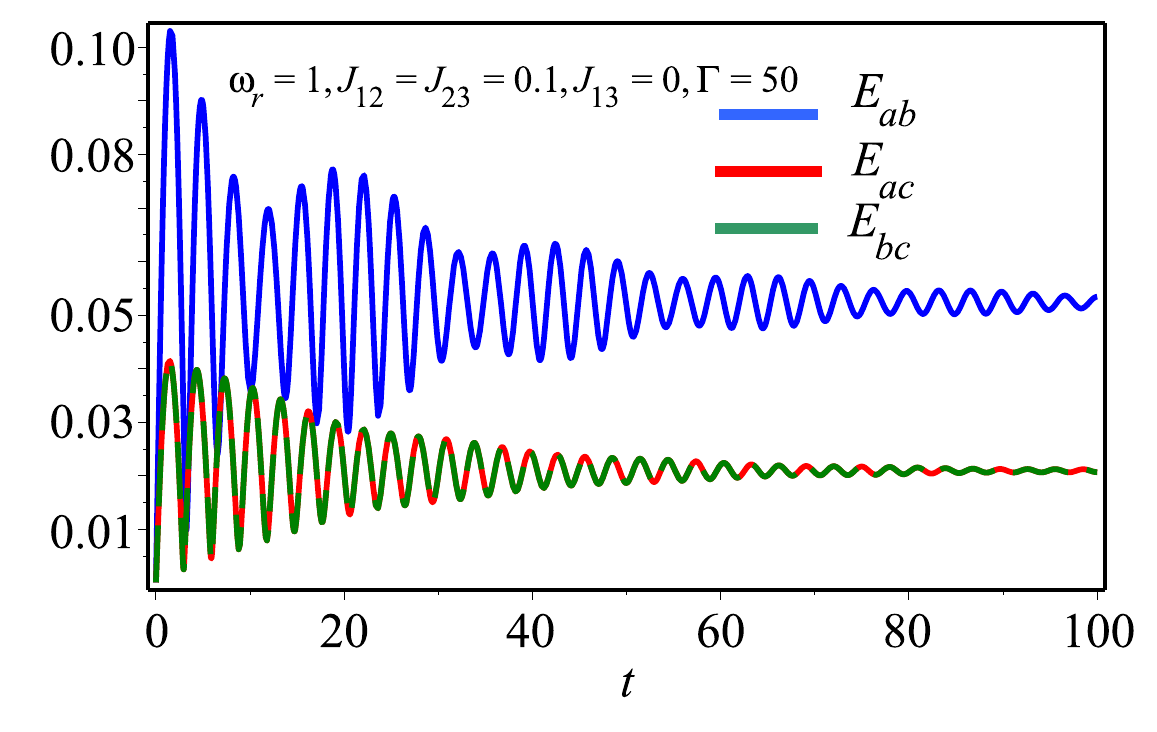}
	\includegraphics[width=8cm, height=5cm]{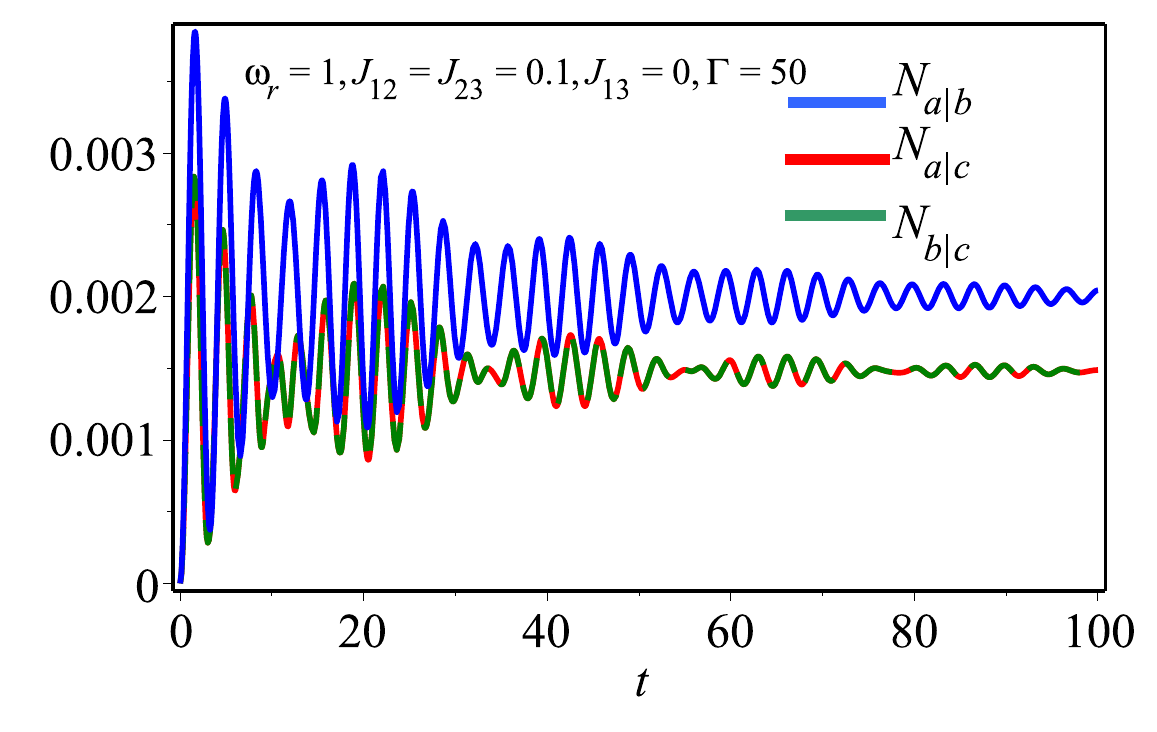}
	\includegraphics[width=8cm, height=5cm]{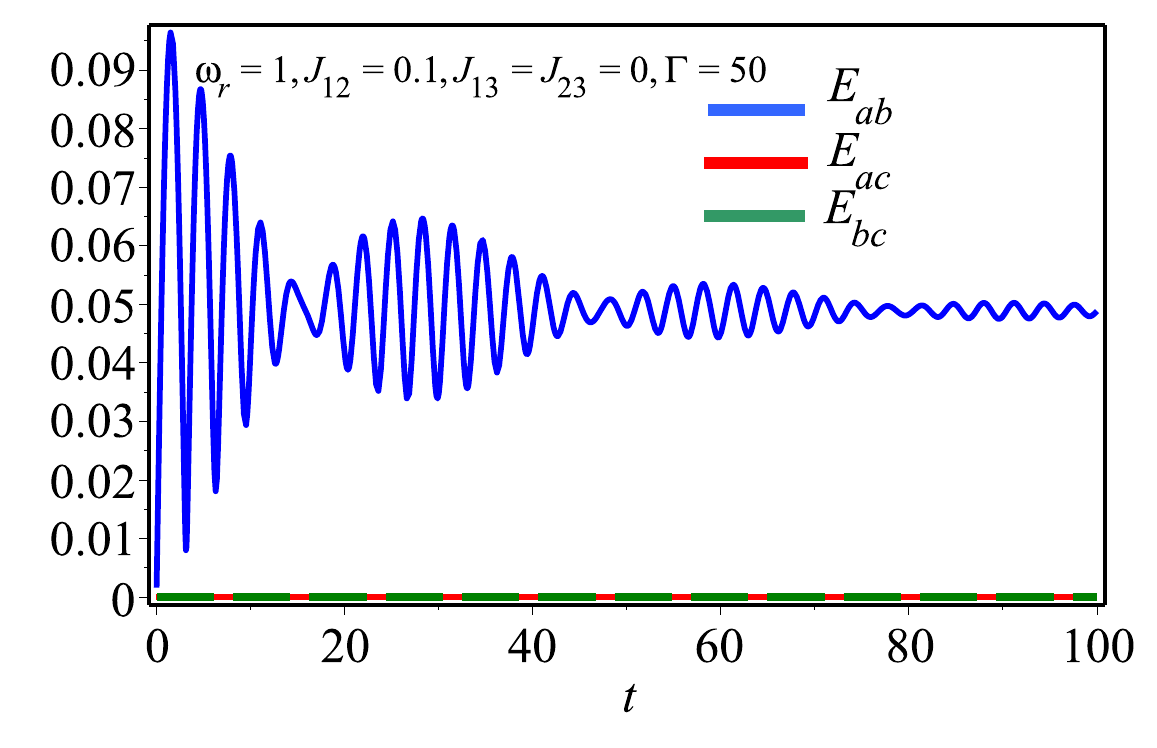}
	\includegraphics[width=8cm, height=5cm]{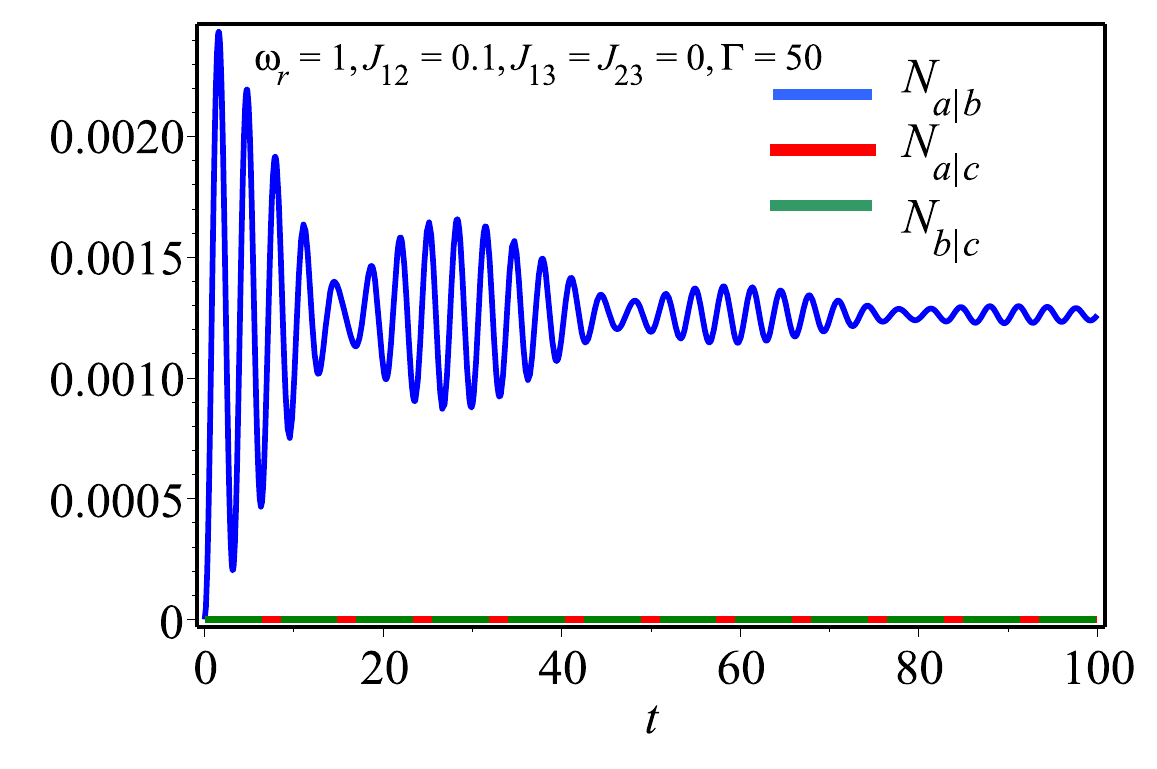}
	
	\caption{(color online) The interplay between the  dynamics of  entanglement and virtual excitations. The parameters $\omega_1=\omega_2=\omega_3=\omega_r=1$, and  $\Gamma=50$ were set.}\label{fig6}
\end{figure}


\section{Concluding remarks}

In this work, we analyzed the interconnection between excitations and entanglement. The system of interest is made up of three non-resonant oscillators. First of all, by making use of Euler rotation and squeezor transformations, we have obtained the diagonalized form of the Hamiltonian in terms of creation and annihilation operators. We have shown that the Milburn density $\rho$ is Gaussian, and this is based on the unitary Milburn evolution, the quadratic form of the Hamiltonian and the Gaussian nature of the initial ground state. Additionally, due to the Gaussian nature of the quantum density, we have derived the Milburn density in its symplectic covariance matrix form. 

 The corresponding covariance matrix is expressed as a Poissonian sum of the product of $26$ symplectic matrices. This finding led us to quantify  the entanglement together with virtual excitations only based on the covariance matrix. Furthermore, we investigated the resonant case and discovered that steady values of excitations are dependent on the coupling $J$ and the resonant frequency $\omega_r$. We have also shown that the necessary time to establish the steady states can be controlled by the  normal frequencies $\Omega_j$ and $\Gamma$. In addition, the excitations were analyzed between all oscillators, and it was found that they exchanged between them. The redistribution of excitations among partitions was investigated, and it was discovered that virtual excitations, like quantum discord, violate the principle of monogamy. And we have shown that excitations are freely generated or annihilated but constrained by a triangular inequality called the polygamy inequality.

As a consequence, excitations are polygamous and are not freely distributed between the parts of the system. This issue can be developed later, in both continuous and discrete variables, to unify and understand the hierarchy of quantum correlations via virtual excitations-based measures. Finally, we have analyzed the interplay between excitations and entanglement. It is found that excitations are necessary to maintain the pairwise entanglement between oscillators. As a result, the extinction of excitations implies disentanglement, thus the hierarchy (in the sense of \cite{hierarchy}) relationship between excitations and entanglement is demonstrated. Another point worth noting is that as one approaches the hermitianity point, i.e., $\min(\Omega_j)\to 0$, the excitations and entanglement become less susceptible to decoherence.

The present work will not remain at this stage, and therefore, we are willing to extend it to deal with other issues of the entanglement and related matters. On the other hand, we hope that our work gives a pulse to a new route to quantifying and processing quantum information.

\end{document}